\documentclass[a4paper,11pt]{article}

\usepackage{fullpage,amsthm,color}
\usepackage{titlesec}

\definecolor{myurlcolor}{rgb}{0,0,0.4}
\definecolor{mycitecolor}{rgb}{0,0.5,0}
\definecolor{myrefcolor}{rgb}{0.5,0,0}
\usepackage{hyperref}
\hypersetup{colorlinks,
linkcolor=myrefcolor,
citecolor=mycitecolor,
urlcolor=myurlcolor}

\usepackage[draft]{fixme}
\usepackage{tikz}
\usetikzlibrary{cd}
\usepackage{subfigure}
\usepackage{multicol}
\usepackage{amsmath,bbm}
\usepackage{graphicx}
\usepackage{amsfonts}
\usepackage{amssymb}
\usepackage{amsmath, amssymb, amsthm,verbatim,graphicx,bbm}
\usepackage{mathrsfs}
\usepackage{color,xcolor,longtable}
\usepackage{changes}
\usepackage{ulem}
\usepackage{adjustbox}

\usetikzlibrary{patterns}
\newcommand{\beq}[0]{\begin{equation}}
\newcommand{\eeq}[0]{\end{equation}}

\newcommand{\op}[1]{#1}				
\newcommand{\avg}[1]{\langle#1\rangle}		
\newcommand{\betaCk}{\beta_C^{k}}
\newcommand{\zetaA}{\zeta_{\mathbf{a}}}
\newcommand{\CA}{\mathcal{C}_{\mathbf{a}}}
\newcommand{\CB}{\mathcal{C}_{\mathbf{b}}}
\newcommand{\CC}{\mathcal{C}_{\mathbf{c}}}
\newcommand{\CBC}{\mathcal{C}_{\mathbf{b}\mathbf{c}}}

\newcommand{\one}{\leavevmode\hbox{\small1\normalsize\kern-.33em1}}
\def\tr{\mbox{tr}}

\def\be{\begin{equation}}
\def\ee{\end{equation}}
\def\ben{\begin{eqnarray}}
\def\een{\end{eqnarray}}
\def\eea{\end{array}}
\def\bea{\begin{array}}

\newcommand{\Tr}[1]{\mathrm{Tr}#1}
\newcommand{\bei}{\begin{itemize}}
\newcommand{\eei}{\end{itemize}}
\newcommand{\ket}[1]{|#1\rangle}

\newcommand{\ketbra}[2]{|{#1}\rangle\!\langle{#2}|}

\renewcommand{\emph}[1]{\textbf{#1}}

\newcommand{\jd}[1]{{\color{cyan}#1}}
\newcommand{\remark}[1]{{\color{magenta}[#1]}}

\theoremstyle{plain}

\theoremstyle{definition}

\theoremstyle{remark}

\titleformat*{\section}{\large\bfseries}

\usepackage{authblk}

\begin{document}

\title{Bell correlations depth in many-body systems}
\author{F. Baccari$^1$, J. Tura$^{1,2}$, M. Fadel$^3$, A. Aloy$^1$, J.-D. Bancal$^{3}$, N. Sangouard$^{3}$, M.~Lewenstein$^{1,4}$, A. Ac\'in$^{1,4}$, and R. Augusiak$^5$}
\affil{$^1$ICFO-Institut de Ciencies Fotoniques, The Barcelona Institute of Science and Technology, 08860 Castelldefels (Barcelona), Spain\\
$^2$Max-Planck-Institut f\"ur Quantenoptik, Hans-Kopfermann-Stra{\ss}e 1, 85748 Garching, Germany
$^3$Department of Physics, University of Basel, Klingelbergstrasse 82, 4056 Basel, Switzerland\\
$^4$ICREA, Pg. Lluis Companys 23, 08010 Barcelona, Spain\\
$^5$Center for Theoretical Physics, Polish Academy of Sciences, Aleja Lotnik\'ow 32/46, 02-668 Warsaw, Poland\\
}
\renewcommand\Affilfont{\itshape\small}

\date{}

\maketitle

\begin{abstract}
While the interest in multipartite nonlocality has grown in recent years, 
its existence in large quantum systems is difficult to confirm experimentally. 
This is mostly due to the inadequacy of standard multipartite Bell inequalities to many-body systems: these inequalities usually rely on expectation values involving many parties and require an individual addressing of each party. In a recent 
work [J. Tura \textit{et al.}, Science \textbf{344}, 6189 (2014)] some of us showed that it is possible to detect nonlocality in multipartite systems using Bell inequalities with only two-body correlators. This opened the way for the detection of Bell correlations with trusted collective measurements through Bell correlation witnesses [R. Schmied \textit{et al.} Science \textbf{352}, 441 (2016)]. These witnesses were recently tested experimentally in many-body systems such as Bose-Einstein condensate or thermal ensembles, hence demonstrating the presence of Bell correlations with assumptions on the statistics. Here, we address the question of assessing the number of particles sharing genuinely nonlocal correlations in a multi-partite system. This endeavor is a priori challenging as known Bell inequalities for genuine nonlocality are based on expectation values involving all the parties, require an individual addressing of parties and a number of measurement settings scaling exponentially with the number of parties. We first show that most of these constraints drop once the witnesses corresponding to these inequalities are expressed: in systems where multipartite expectation values can be evaluated, these witnesses can reveal genuine nonlocality for an arbitrary number of particles with just two collective measurements. We then focus on two-body Bell inequalities and show that they not only allow one to reveal nonlocality in a multipartite quantum state, but also provide information about the number of particles that are genuinely nonlocal. To this end, we introduce a general framework 
allowing to address this problem. Then, we characterize all Bell-like inequalities for a finite number of parties based on symmetric two-body correlations, and provide witnesses that can be used in present day experiments  to reveal a Bell correlation depth $k\leq6$ for any number of parties. 
A violation of our witness for depth $6$ is achieved with existing data from an ensemble of 480 atoms.
\end{abstract}

\section{Introduction}

Local measurements on composite quantum systems may lead to fascinating correlations that cannot be explained by any local realistic theory such as classical physics. This phenomenon, usually referred to as nonlocality (cf. Ref. \cite{Review1}), proves that 
the laws of physics at the micro-scale significantly differ from those at the macro-scale. More importantly, in recent years it has been understood that nonlocality is a powerful resource for device-independent applications that are not achievable within classical physics, with the most prominent examples being device-independent quantum key distribution \cite{Ekert,DIKey}, device-independent entanglement detection \cite{BancalDIEW},
generation and amplification of randomness \cite{PironioRand,ColbeckPhD,RennerColbeck}, or self-testing \cite{MayersYao}.

However, to be able to fully exploit nonlocality as a resource one first needs efficient methods to detect the composite quantum systems that can exhibit it. Since these systems can produce nonlocality upon measurement, i.e. statistics violating a Bell inequality~\cite{Bell}, we say that their state is Bell correlated. Bell inequalities are naturally the most common way of revealing both nonlocal statistics and Bell correlated states.
These are inequalities constraining the set of local realistic correlations, and their violation signals nonlocality. Considerable amount of effort has been devoted to introduce various constructions of Bell inequalities \cite{MultipBell}. Still, the problem of nonlocality detection is much less advanced in the multipartite case than in the bipartite one, in which recently loophole-free Bell tests have been performed \cite{loophole1,loophole2}. There are two main reasons for that: (i) the mathematical complexity of finding all Bell inequalities grows exponentially with the number of parties, (ii) experimental verification of nonlocality is much more demanding in the multipartite case; in particular, individual settings assignment for each party is needed to test a Bell inequality, and most of the known multipartite Bell inequalities (such as for instance the Mermin Bell inequality) involve a large number of measurements, and are constructed from correlation functions involving measurements by all parties (see Ref. \cite{MultipBell}). Therefore, such inequalities are not suited to detect nonlocal correlations in many-body systems in which only a few collective measurements can be applied and one typically has access only to two-body correlations.

One of the ways to tackle these difficulties in the multipartite case, recently put forward in Ref. \cite{ScienceOur,Schmied}, is to use a witness of Bell correlations constructed from a Bell inequality involving only two-body correlators. On one hand, this reduces the mathematical complexity of finding the Bell inequality as the dimension of the Bell polytope is much smaller than in the general case. On the other hand, the resulting witnesses can be expressed in terms of just a small number of collective two-body expectation values, which are routinely measured in certain many-body quantum systems (see Refs. \cite{SpinPolSp,Eckert}). This makes these witnesses very practical. Recently, in a series of 
papers \cite{Series1,Series2,Series3,BaselMany} some of us have demonstrated that Bell inequalities involving only two-body correlations are powerful enough to reveal nonlocality in composite quantum systems with an arbitrary number of subsystems. The corresponding witnesses were tested in two experiments recently reporting on Bell correlations in gaussian many-body states consisting of 480 atoms \cite{Schmied} in a Bose-Einstein condensate and $5\cdot 10^5$ particles \cite{Engelsen} in a thermal ensemble. 

However, violation of these Bell inequalities and the corresponding witnesses only signal the presence of some kind of Bell correlations. In fact, they are unable to provide information regarding how many particles genuinely share Bell correlations. This naturally raises the question on how to reveal the depth of non-locality in many-body systems. At first sight, the problem is challenging. Known Bell inequalities for genuine non-locality not only use expectation values involving all parties, they also require the ability to perform a different measurement on each party. Furthermore, the number of measurement settings scales exponentially with the number of parties. 

The main aim of this work is to address this problem. Here, we first examine known Bell inequalities for genuine nonlocality in the context of many-body systems. We show that the corresponding witnesses use expectation values involving all the parties but that they can be expressed with two collective measurements only. This gives access to genuine Bell correlations in many-body systems where one high-order measurement can be performed. We then turn to the problem of revealing the non-locality depth in multi-partite systems using two-body correlations only, hence guaranteeing that no high-order moment will be necessary at the level of the witness. For this, we introduce a general framework. The problem of detection of genuine nonlocality in this context is characterized for a relatively small number of  parties, providing lists of Bell-like inequalities that do the job. Moreover, we give a Bell-like inequality detecting the nonlocality depth from 1 to 7 for any number of parties. This allows us to derive witnesses using two-body correlations and collective measurements to test the non-locality depth of a Bose-Einstein condensate with 480 atoms. 

The structure of this work is as follows. Sec. \ref{Preliminaries} introduces all concepts relevant for upcoming sections such as a Bell scenario and correlations, the notions of $k$-producibility of nonlocality and nonlocality depth. In Sec. \ref{Sec2bis}, we derive from a known family of Svetlichny inequalities a witness for revealing genuine non-locality with two collective measurements, including one many-body expectation value. In Sec. \ref{Sec3}, we discuss how to describe $k$-producible correlations in terms of the symmetric two-body correlations. Then, in Sec. \ref{Sec4}, we present our main result, i.e., Bell-like inequalities detecting $k$-nonlocality for a few values of $k$ and any number of parties $N$ while in Sec. \ref{Sec5}, we focus on the corresponding witness. We conclude in Sec. \ref{Conclusion} in which we also outline possible directions for further research.

\section{Preliminaries}
\label{Preliminaries}

Before passing to our results let us set some background information
concerning the Bell scenario and correlations, and also introduce such notions
as $k$-producility of nonlocality, nonlocality depth and genuine nonlocality in 
the multipartite scenario.

\subsection{Bell scenario and correlations}

Consider $N$ parties $A_1,\ldots,A_N$ sharing some $N$-partite physical system. Each party is allowed to perform measurements on their share of this system. 
Here we consider the simplest scenario in which party $A_i$ 
has two observables $M_{x_i}^{(i)}$ with $x_i=0,1$ to choose from, 
and each observable is assumed to yield two outcomes
$a_i=\pm 1$. Such a scenario is often referred to as the $(N,2,2)$ scenario. 

Such local measurements lead to correlations that are described by the following set of conditional probabilities 
\begin{equation}\label{correlations0}
\{p(a_1,\ldots,a_N|x_1,\ldots,x_N)\},
\end{equation}
where each $p(\mathbf{a}|\mathbf{x})=p(a_1,\ldots,a_N|x_1,\ldots,x_N)$ stands
for the probability that the parties obtain outcomes $a_1,\ldots,a_N=:\mathbf{a}$
upon perfoming the measurements labelled by $x_1,\ldots,x_N=:\mathbf{x}$. Clearly, 
these probabilities are non-negative, i.e., 
\begin{equation}\label{posit}
p(\mathbf{a}|\mathbf{x})\geq 0
\end{equation}
holds for any $\mathbf{a}$ and $\mathbf{x}$, and are normalized, that is, 
\begin{equation}\label{normalization}
\sum_{\mathbf{a}}p(\mathbf{a}|\mathbf{x})=1
\end{equation}
for any $\mathbf{x}$. Naturally, they are also assumed to satisfy the no-signaling 
principle in a space-like setting, which says that faster than light communication between 
various parties is not possible. In mathematical terms this is equivalent to the following set of linear constraints
\begin{equation}\label{nonsignaling}
\sum_{a_i}\left[p(a_1,\ldots,a_i,\ldots,a_N|x_1,\ldots,x_i,\ldots,x_N)-p(a_1,\ldots,a_i,\ldots,a_N|x_1,\ldots,x_i',\ldots,x_N)\right]=0
\end{equation}
for all $x_i,x_i'$, $a_1,\ldots,a_{i-1},a_{i+1},\ldots,a_N$ and $x_{1},\ldots,x_{i-1},x_{i+1},\ldots,x_N$ and all $i=1,\ldots,N$.

Now, given the definition of nonsignaling correlations, it is useful to notice that in the simplest scenario of each party performing dichotomic measurements, they can be equivalently described by a collection of expectation values
\begin{equation}\label{correlators}
\left\{\left\langle M^{(i_1)}_{x_{i_1}}\ldots M^{(i_k)}_{x_{i_k}}\right\rangle\right\}_{i_1,\ldots,i_k;x_{i_1},\ldots,x_{i_k};k},
\end{equation}
with $i_1<\ldots <i_k=1,\ldots,N$ and $k=1,\ldots,N$ (all possible expectation values, ranging from the single-body to $N$-partite ones are taken into account). 
These two representations of correlations are related through the formula
\begin{equation}\label{relation}
p(\mathbf{a}|\mathbf{x})=\frac{1}{2^N}\left[1+\sum_{k=1}^N\sum_{1\leq i_1<i_2<\ldots<i_k\leq N}a_{i_1}\ldots a_{i_k}\left\langle M^{(i_1)}_{x_{i_1}}\ldots M^{(i_k)}_{x_{i_k}}\right\rangle\right],
\end{equation}
that holds for any $\mathbf{a},\mathbf{x}$. Note that via (\ref{posit}) this imposes a set of linear constraints on the values of correlators in (\ref{correlators}) in any theory obeying the no-signaling principle. It also follows that in the $(N,2,2)$ scenario 
the set (\ref{correlators}) contains $D(N)=3^N-1$ expectation values and we can conveniently think of them as a vector living in the real space $\mathbbm{R}^{D(N)}$. 
It should be noticed in passing that in the present work most of our results are stated in terms of expectation values (\ref{correlators}). Still, some of the definitions such as those of local correlations or $k$-producible correlations are for simplicity stated in the probability picture. It is, however, straightforward to reformulate them in terms of correlators by using the Fourier transform (\ref{relation}).

Imagine now that correlations (\ref{correlations0}), or, equivalently, (\ref{correlators}) arise by performing local measurements on some quantum state $\rho_N$. 
In such a case each expectation value (\ref{correlators}) can be expressed \textit{via} Born's rule as
\begin{equation}
\left\langle M^{(i_1)}_{x_{i_1}}\ldots M^{(i_k)}_{x_{i_k}}\right\rangle=\Tr\left[\rho_{A_{i_1}\ldots A_{i_k}}\left(M^{(i_1)}_{x_{i_1}}\otimes \ldots \otimes  M^{(i_k)}_{x_{i_k}}\right)
\right],
\end{equation}
where $\rho_{A_{i_1}\ldots A_{i_k}}$ stands for a subsystem of 
$\rho_N$ corresponding to the parties $A_{i_1},\ldots,A_{i_k}$, whereas  $M^{(i)}_{x_{i}}$ denote now the standard quantum observables, i.e., Hermitian operators with eigenvalues $\pm 1$. 
It turns out that the set of all correlations obtained from quantum states in the above experiment, denoted $\mathcal{Q}_N$, is convex if no assumption on the local dimension of the state used in the experiment is made. It is a superset for another relevant set of correlations, which consists of those correlations that can be obtained by the parties when the only resource they share is classical information represented by some random variable 
$\lambda$ with a probability distribution $p(\lambda)$. The most general form of such correlations is (cf. also Refs. \cite{Review1,Review2}) 
\begin{equation}\label{LHV}
\left\langle M^{(i_1)}_{x_{i_1}}\ldots M^{(i_k)}_{x_{i_k}}\right\rangle=\sum_{\lambda}p(\lambda)\left\langle M^{(i_1)}_{x_{i_1}}\right\rangle_{\lambda}\cdot\ldots\cdot \left\langle M^{(i_k)}_{x_{i_k}}\right\rangle_{\lambda},
\end{equation}
where $\langle M^{(i)}_{x_{i}}\rangle_{\lambda}\in[-1,1]$ are single-party expectation values enumerated by $\lambda$ for any $x_i$ and $i$. 
It follows that the set of local correlations is a polytope
in $\mathbbm{R}^{D(N)}$ which we denote $\mathcal{P}_N$. Recall that a polytope is the convex hull of a finite number of vertices (see also Appendix \ref{AppA} for a more formal definition). Its extremal points (vertices) are those correlations for which all expectation values (\ref{correlators}) are product and the local single-party mean values equal $\pm 1$. In other words, any vertex of $\mathcal{P}_N$ represents correlations that the parties can produce by using local deterministic strategies, i.e., 
each local measurement has a perfectly determined outcome. In this scenario, the number of vertices of $\mathcal{P}_N$ is $2^{2N}$. 

Clearly, $\mathcal{P}_N\subset \mathcal{Q}_N$, and Bell showed \cite{Bell} that this inclusion is strict, that is, there are quantum correlations that do not admit a decomposition of the form (\ref{LHV}). To this aim he used linear inequalities, usually termed Bell inequalities, that constrain the local 
set $\mathcal{P}_N$. Violation of a Bell inequality implies that given correlations cannot be expressed in the form (\ref{LHV}), in which case we call such correlations nonlocal.

For further benefits let us also notice that 
the nonsignaling correlations, i.e., those sets (\ref{correlations0}) that satisfy 
conditions (\ref{posit}), (\ref{normalization}) and the linear constraints 
(\ref{nonsignaling}), form a polytope too, denoted $\mathcal{NS}_N$. Contrary to the local polytope $\mathcal{P}_N$, it has a very simple
dual description in terms of facets: all its facets correspond simply to the non-negativity constraints (\ref{posit}) for all $N$-tuples $\mathbf{a}$ and $\mathbf{x}$. On the other hand, it is generally an extremely difficult problem to determine the vertices of $\mathcal{NS}_N$; they were found only in the simplest scenarios: (i) two observers, each performing two measurements with any number of outcomes \cite{Barrett}, (ii) two observers perfoming any number of dichotomic measurements \cite{Lluis}, and (iii) three parties, each performing two dichotomic measurements \cite{NS}. It should also be mentioned that in the $(N,2,2)$ scenario some partial progress in this direction was made in Ref. \cite{Fritz}. Concretely, by establishing the duality between Bell inequalities and the vertices of the nonsignaling polytope in this scenario, some of its vertices were identified: those that correspond to tight Bell inequalities such as the correlation inequalities found in Refs. \cite{MultipBell}.

Let us conclude by noting that in general $\mathcal{P}_N \subset \mathcal{Q}_N \subset\mathcal{NS}_N$ and both inclusions are strict; the first one follows from the work of Bell, while the second one is due to the fact that there exist correlations that despite satisfying the no-signaling principle do not have quantum representation \cite{PRbox}.

\subsection{The concepts of $k$-producibility of nonlocality and nonlocality depth}
\label{k-producibility}

As already said, violation of Bell inequalities signals nonlocality, however, in the multipartite scenario it tells us nothing about how many parties actually share nonlocal correlations; to give an illustrative example, a state like $(1/\sqrt{2})(\ket{00}+\ket{11})\otimes\ket{0}$ that is the product of the maximally entangled state of two qubits
with some additional state and the GHZ state $(1/\sqrt{2})(\ket{000}+\ket{111})$ are both nonlocal, however, the first one has only two-body nonlocality, while the second one is nonlocal "everywhere".

Several approaches have been proposed to describe the types of nonlocality that can appear in the multipartite scenario~\cite{MultipartiteNonloc1,MultipartiteNonloc2}. Following~\cite{MultipartiteNonloc2}, we choose here the notion of $k$-producibility of nonlocality or nonlocality depth, which goes along the lines developed to describe mutipartite entanglement (see Refs. \cite{Otfried1}). To this end, we partition the set $I=\{1,\ldots,N\}$ into $L$ pairwise disjoint non-empty subsets $\mathcal{A}_i$ such that by joining them one recovers $I$ and the size of each $\mathcal{A}_i$ is at most $k$ parties. We call such a partition $L_k$-partition of $I$. We call correlations $\{p(\mathbf{a}|\mathbf{x})\}$ $k$\textit{-producible with respect to the given $L_k$-partition} if they admit the following decomposition
\begin{equation}\label{knonloc}
p(\mathbf{a}|\mathbf{x})=\sum_{\lambda}p(\lambda)p_1(\mathbf{a}_{\mathcal{A}_1}|\mathbf{x}_{\mathcal{A}_1},\lambda)\cdot\ldots\cdot p_L(\mathbf{a}_{\mathcal{A}_L}|\mathbf{x}_{\mathcal{A}_L},\lambda)
\end{equation}
where $\mathbf{a}_{\mathcal{A}_i}$ and $\mathbf{x}_{\mathcal{A}_i}$ are outcomes and measurements choices corresponding to the observers belonging to $\mathcal{A}_i$.
It is known that the probability distribution $p_i(\mathbf{a}_{\mathcal{A}_i}|\mathbf{x}_{\mathcal{A}_i},\lambda)$ here cannot be left unconstrained, but should be chosen to correspond to a given resource $\mathcal{R}$ in order to avoid describing a self-contradicting model~\cite{Definition1,Definition2}. In our case, we choose $\mathcal{R}=\mathcal{NS}$, i.e. all conditional probabilities in Eq.~\eqref{knonloc} are required to satisfy the no-signaling condition~\eqref{nonsignaling}. This solves this issue and leads to the identification of $k$-way NS nonlocal correlations.
One could also make a more restrictive assumption that all $\{p_i(\mathbf{a}_{\mathcal{A}_i}|\mathbf{x}_{\mathcal{A}_i},\lambda)\}$ are quantum correlations; this, however, will be explored in another publication \cite{OurQuantum}.

We then call correlations $\{p(\mathbf{a}|\mathbf{x})\}$
$k$\textit{-producible} if they can be written as a convex combination of correlations
that are $k$-producible with respect to different $L_k$-partitions, i.e., 
\begin{equation}\label{kproducibility}
p(\mathbf{a}|\mathbf{x})=\sum_{S\in S_k}q_S p_S(\mathbf{a}|\mathbf{x})
\end{equation}
where $S_k$ is the set of all $L_k$-partitions and $p_S(\mathbf{a}|\mathbf{x})$ are correlations that admit the decomposition (\ref{knonloc}) with respect to the $k$-partition $S$. The minimal $k$ for which the correlations $\{p(\mathbf{a}|\mathbf{x})\}$ are of the form (\ref{kproducibility}) is called \textit{nonlocality depth}\footnote{A more adequate terminology  would be \textit{nonlocality depth with respect to nonsignaling resources} or simply \textit{NS nonlocality depth} as in the definition of $k$-producibility we assume the probabilities $p_i(\mathbf{a}_{\mathcal{A}_i}|\mathbf{x}_{\mathcal{A}_i},\lambda)$ to be nonsignaling, whereas according to Refs. \cite{Definition1,Definition2} other types of resources can also be considered.}. Correlations whose nonlocality depth is $k$ are \textit{genuinely $k$-partite nonlocal} or \textit{simply $k$-nonlocal} as
there must exist a subset of $k$ parties which share nonsignaling correlations that are genuinely nonlocal~\cite{Svetlichny87}.

Let us notice that in the particular case of $k=1$ there is only one, up to permutations, $L_1$-partition: each party forms a singleton $\mathcal{A}_{i} =  \{A_i\}$ $(i=1,\ldots,N)$. In this case the above definition recovers the above introduced definition of fully local  correlations (\ref{LHV}). 
Then, on the other extreme of $k=N$, we have correlations in which all parties share nonlocality and are thus called genuinely multiparty nonlocal (GMNL).

Geometrically, as in the case of fully local models (\ref{LHV}), $k$-producible correlations form polytopes, denoted $\mathcal{P}_{N,k}$ (we use this notation for both the probability and correlators picture). Vertices of these polytopes are 
product probability distributions of the form 
\begin{equation}\label{vertex}
p(\mathbf{a}|\mathbf{x})=p_1(\mathbf{a}_{\mathcal{A}_1}|\mathbf{x}_{\mathcal{A}_1})\cdot\ldots\cdot p_L(\mathbf{a}_{\mathcal{A}_L}|\mathbf{x}_{\mathcal{A}_L})
\end{equation}
with each $p_i(\mathbf{a}_{\mathcal{A}_i}|\mathbf{x}_{\mathcal{A}_i})$ being a vertex of the corresponding $|\mathcal{A}_i|$-partite nonsignaling polytope with $|\mathcal{A}_i|\leq k$ for all $i$; if for some $i$, $|\mathcal{A}_i|=1$, then
$p_i(\mathbf{a}_{\mathcal{A}_i}|\mathbf{x}_{\mathcal{A}_i})\equiv p_i(a_i|x_i)$ is simply a deterministic probability distribution, i.e., $p_i(a_i|x_i)\in\{0,1\}$ for all values of $x_i$. One then needs to consider all $L_k$-partitions in order to construct all vertices of $\mathcal{P}_{N,k}$. It thus follows that to construct all vertices of $\mathcal{P}_{N,k}$ one needs to know the vertices of the $p$-partite nonsignaling polytopes for all $p\leq k$.

Let us finally notice that with the aid of the formula (\ref{relation}) all the above definitions, in particular the one of $k$-producible correlations as well as the vertices of $\mathcal{P}_{N,k}$ can be equivalently formulated in terms of correlators (\ref{correlators}); in particular, for a vertex (\ref{vertex}) the correlators (\ref{correlators}) factorize whenever the parties belong to different groups $\mathcal{A}_i$. 

Thus, in order to detect nonlocality depth of given correlations one can follow the standard approach used to reveal nonlocality: construct Bell-like inequalities that constrain 
$\mathcal{P}_{N,k}$, that is, inequalities of the following form
\begin{equation}\label{BellLikeIn}
\sum_{\mathbf{a},\mathbf{x}}T_{\mathbf{a},\mathbf{x}}\,p(\mathbf{a}|\mathbf{x})\leq \beta_{C}^k,
\end{equation}
where $\beta_{C}^k$ is the maximal value of the Bell expression $\sum_{\mathbf{a},\mathbf{x}}T_{\mathbf{a},\mathbf{x}}\,p(\mathbf{a}|\mathbf{x})$ over all correlations belonging to $\mathcal{P}_{N,k}$; in the particular case $k=1$, $\beta_C^1$ is the local or classical bound the Bell inequality (\ref{BellLikeIn}). Ideally, Bell-like inequalities that correspond to facets of $\mathcal{P}_{N,k}$ would be the strongest tests of nonlocality depth. Violation of a Bell-like inequality (\ref{BellLikeIn}) implies that $\{p(\mathbf{a}|\mathbf{x})\}$ is at least genuinely $k+1$-nonlocal, or, in other words, that these correlations have nonlocality depth at least $k+1$. Let us finally mention that the maximal value of the Bell expression (\ref{BellLikeIn}) over all quantum correlations will be denoted $\beta_Q$.

\section{Witnessing genuine non-locality from Svetlichny and Mermin inequalities}
\label{Sec2bis}
The Mermin and Svetlichny Bell expressions are known to be suitable for the detection of non-locality depth in multi-partite systems~\cite{MultipartiteNonloc1}. They are thus a natural starting point for our investigations. We here show the form of the corresponding witnesses for non-locality depth.
  
\subsection{Bell operators for Svetlichny and Mermin inequalities}

Let us begin with the Svetlichny Bell expressions 
written in the following form~\cite{Barreiro13}
\begin{equation}
\label{In_Svetlichny}
I_N^\text{Svet} = 2^{-N/2}\left[\sum_{\mathbf{x}|\mathbf{s} = 0\ \text{(mod 2)}}(-1)^{\mathbf{s}/2}E_{\mathbf{x}}+\sum_{\mathbf{x}|\mathbf{s} = 1\ \text{(mod 2)}}(-1)^{(\mathbf{s}-1)/2}E_{\mathbf{x}}\right],
\end{equation}
where $\mathbf{s}=\sum_i x_i$ is the sum of all parties' settings (recall that $x_i\in\{0,1\}$), $\mathbf{x}|\mathbf{s} = i\ \text{(mod 2)}$ means that the summation is over those $\mathbf{x}$'s for which $\mathbf{s}$ is even for $i=0$
or odd for $i=1$, and, finally, 
\begin{equation}\label{Npartite}
E_{\mathbf{x}} = \left\langle M^{(1)}_{x_{1}}\ldots M^{(N)}_{x_{N}}\right\rangle
\end{equation}
is a short-hand notation for an $N$-partite correlator. The fully local $\beta_C^1$, the $k$-nonlocal $\beta_C\equiv \beta_C^k$, the quantum $\beta_Q$ and the nonsignaling bounds are given in Table \ref{table_bounds_Svetlichny}.
\begin{table}[h!]
\centering
\begin{tabular}{ccccc}
& local $\beta_C$ & $k$-nonlocal $\beta_C^k$ & quantum $\beta_Q$ & nonsignaling $\beta_{NS}$  \\
\hline 
   & $2^{\frac{1-(-1)^N}4}$  & $2^{(N-2\lfloor\frac{\lceil N/k\rceil}{2}\rfloor)/2}$ & $2^{(N-1)/2}$ & $2^{N/2}$ \\
\hline
\end{tabular}
\caption{The local $\beta_C$, $k$-nonlocal $\beta_C^k$, quantum $\beta_Q$ and nonsignaling bounds for the Svetlichny Bell expression.}
\label{table_bounds_Svetlichny}
\end{table}
Generally, the Svetlichny-type bounds for these inequalities are expressed in terms of the number of groups $m$ in which the $N$ parties are splitted. Noticing that $N$, $m$ and $k$ are related by the relation $m+k-1 \leq N \leq m k$ allows one to express the bound as a function of the nonlocality depth $k$ (resulting in the bound in the table above). The fact that the Svetlichny bounds can be achieved with a model in which $\lfloor N/k\rfloor$ groups contain exactly $k$ parties implies that the resulting bounds are tight~\cite{MultipartiteNonloc1}.

As the quantum bound is larger than the $(N-1)$-nonlocal bound, the Svetlichny expression can reveal genuine nonlocality. The two sums in Eq. \eqref{In_Svetlichny}, however, involved $2^N$ terms in total. This makes the Svetlichny inequality difficult to test in systems with a large number of parties. But we are not interested in performing a Bell test. Rather, we are willing to assume that the measurements are well calibrated spin projections. We thus proceed to derive the device dependent witness corresponding to these inequalities. For the measurement settings maximizing the quantum value for a $|\mathrm{GHZ}_N^+\rangle$ state, $\ket{\rm{GHZ}_N^\pm}=(\ket{0}^{\otimes N}\pm\ket{1}^{\otimes N})/\sqrt{2},$ that is the measurements given by 
\begin{equation}
M^{(i)}_j=\cos(\phi_j)\,\sigma_x + \sin(\phi_j)\,\sigma_y\ ,\qquad \phi_j=-\frac{\pi}{4N} + j\frac{\pi}{2}
\end{equation}
with  $j\in 0,1,$ we find that the Svetlichny operator takes the following very simple form
\begin{equation}
\mathcal{B}_N^\text{Svet} = 2^{(N-1)/2}\left(\ketbra{0}{1}^{\otimes N}+\ketbra{1}{0}^{\otimes N}\right).
\end{equation}
If the mean value of this operator $\tr(\rho \mathcal{B}_N^\text{Svet})$ is larger than the $k$-nonlocal bound given in the table above, we conclude that the state $\rho$ has the capability to violate a Svetlichny inequality with the corresponding bound, that is, $\rho$ is $(k+1)$-Bell correlated.

Before showing how to access the mean value of $\mathcal{B}_N^\text{Svet},$ let us do a similar work for the Bell expression corresponding to the Mermin inequality~\cite{Mermin90}.
Let us consider the following form of the Mermin Bell expression
\begin{equation}\label{Mermin}
I_N^\text{Mermin} = 2^{-(N-1)/2}\left[\sum_{\mathbf{x}|\mathbf{s} = 0\ \text{(mod 2)}}(-1)^{\mathbf{s}/2}E_{\mathbf{x}}\right],
\end{equation}
where, as before, $\mathbf{s}=\sum_i x_i$ and $E_{\mathbf{x}}$ is given by Eq. (\ref{Npartite}). The corresponding classical, quantum, $k$-nonlocal and nonsignaling bounds are given in Table \ref{table_Mermin}.
\begin{table}[h!]
\centering
\begin{tabular}{ccccc}
& local $\beta_C$ & $k$-nonlocal $\beta_C^k$ & quantum $\beta_Q $ & nonsignaling $\beta_{NS}$ \\
\hline
   & $2^{\frac{1+(-1)^N}4}$  & $2^{(N-2\lfloor\frac{\lceil N/k \rceil+1}{2}\rfloor+1)/2}$ & $2^{(N-1)/2}$ & $2^{(N-1)/2}$ \\
\hline
\end{tabular}
\caption{The local $\beta_C$, $k$-nonlocal $\beta_C^k$, quantum $\beta_Q$ and nonsignaling bounds for the Mermin Bell expression.}
\label{table_Mermin}
\end{table}
The settings leading to the maximal quantum violation of the Mermin Bell inequality for the $|\mathrm{GHZ}^+_N\rangle$ state are the same for each party and are given by $M_0^{(i)}=\sigma_x$ and $M_1^{(i)}=\sigma_y$, and, interestingly, for them the Bell operator reduces
to the same Bell operator as before, that is,
\begin{equation}
\label{mermin_op}
\mathcal{B}_N^\text{Mermin} = 2^{(N-1)/2}\left(\ketbra{0}{1}^{\otimes N}+\ketbra{1}{0}^{\otimes N}\right).
\end{equation}
In other words, for the optimal choices of observables, the same Bell operator can be attributed to Svetlichny and Mermin Bell expression. We can thus derive a common witness with which the nonlocality depth of any multipartite system can be evaluated. We elaborate on this in the upcoming section.

\subsection{Witnessing genuine nonlocal correlations with 2 measurements}
In order to state our bound we first need to prove that the following operator
\begin{equation}
\nonumber
\chi_N = |\text{GHZ}^+_N\rangle\! \langle \text{GHZ}^+_N | - |\text{GHZ}^-_N\rangle\! \langle \text{GHZ}^-_N| - \underbrace{\sigma_x^1 \hdots \sigma_x^N}_{N\,\text{times}} - \sum_{m\neq n}^N \sigma_z^{m}\sigma_z^n + N(N-1)\mathbbm{1}
\end{equation}
is positive semi-definite, where $\sigma_{x/z}^i$ stands for the Pauli matrix $\sigma_{x/z}$ acting on site $i$. The proof can be derived in various ways from \cite{Toth05}, this reference focusing on genuine entanglement detection. 

To this aim, let us assume for simplicity $N$ to be even and consider 
the GHZ state $|\text{GHZ}^{+}_N\rangle,$ and the following set of states 
obtained by flipping $k$ of its spins with $k=1,\ldots,N/2$, that is
\begin{eqnarray}\label{states}
\sigma_x^{i_1} |\text{GHZ}^{+}\rangle,&\quad& \nonumber\\
\sigma_x^{i_1}\sigma_x^{i_2} |\text{GHZ}^{+}\rangle,&\quad& i_1\neq i_2\nonumber\\
&\vdots& \nonumber\\
\sigma_x^{i_1}\sigma_x^{i_2}\ldots\sigma_{x}^{i_{N/2-1}} |\text{GHZ}^{+}\rangle,&\quad& i_1\neq  i_2\neq \ldots\neq i_{N/2-1}\nonumber\\
\sigma_x^{i_1}\sigma_x^{i_2}\ldots\sigma_{x}^{i_{N/2}} |\text{GHZ}^{+}\rangle,&\quad& i_1\neq  i_2\neq \ldots\neq i_{N/2}
\end{eqnarray}
where $i_{\ell}=1,\ldots,N$ for $\ell=1,\ldots,N/2$. Notice that in each "line" of
Eq. (\ref{states}) there are $C_N^{k}=\binom{N}{k}$ $(k=1,\ldots,N/2-1)$ 
orthogonal states except for the last one in which the number of orthogonal vectors is $C_N^{N/2}/2$. We then construct an analogous set of vectors with $|\mathrm{GHZ}_N^-\rangle$, 
which altogether gives us a set of 
$$
2 \sum_{k=0}^{N/2-1} C_N^k + C_N^{N/2} = \sum_{k=0}^{N} C_N^k = 2^N
 $$
orthonormal vectors forming a basis in $(\mathbbm{C}^2)^{\otimes N}$. Let us now show that
the operator $\chi_N$ is diagonal in this basis. For this purpose, we notice that 
$\sigma_x \sigma_z \sigma_x=-\sigma_z$ and therefore (see also Ref. \cite{ScaraniGisin})
\begin{eqnarray}
\nonumber
&&\langle \text{GHZ}_N^\pm |( \sigma_x^{i_1} \sigma_x^{i_2} \hdots \sigma_x^{i_{\ell}})\, (\sigma_z^{m} \sigma_z^n)\,( \sigma_x^{i_1} \sigma_x^{i_2} \hdots \sigma_x^{i_{\ell}} )|\text{GHZ}_N^\pm \rangle\\
\nonumber
&& = (-1)^{\lambda_{m,n}} \tr\left(\sigma_z^{m} \sigma_z^n |\text{GHZ}_N^\pm \rangle\! \langle \text{GHZ}_N^\pm |\right) = (-1)^{\lambda_{m,n}},
\end{eqnarray}
where $\ell=1,\ldots,N/2$, $m\neq n$, and $\lambda_{m,n}=0$ if both qubits $m$ and $n$ are flipped or neither of them, and $\lambda_{m,n}=-1$ if only one of them is flipped. We also notice that for the parity operator one has
\begin{equation}
\nonumber
\langle \text{GHZ}_N^\pm | (\sigma_x^{i_1} \sigma_x^{i_2} \hdots \sigma_x^{i_{\ell}}) \,(\sigma_x^{1} \hdots \sigma_x^N)\,( \sigma_x^{i_1} \sigma_x^{i_2} \hdots \sigma_x^{i_{\ell}} )|\text{GHZ}_N^\pm \rangle = \pm 1.
\end{equation}

All this means that the operator $\chi_N$ is diagonal in the above basis. Furthermore, the maximal eigenvalue of $\sigma_x^1 \hdots \sigma_x^N + \sum_{m\neq n}^N \sigma_z^{m}\sigma_z^n$ is  $N(N-1)+1$ and the corresponding eigenstate is $|\text{GHZ}^+_N\rangle$. Then, the other GHZ state $|\text{GHZ}^-_N\rangle$ corresponds to the eigenvalue $N(N-1)-1$ and all the other elements of the above basis vectors eigenvectors with eigenvalues smaller or equal to $N(N-1)-1$. As a result, all eigenvalues of $\chi_N$ are non-negative, and hence
\begin{eqnarray}
\nonumber
\mathcal{B}_N^\text{Svet}=\mathcal{B}_N^\text{Mermin}&\!\!\!=\!\!\!&\sqrt{2}^{N-1}\left(|\text{GHZ}_N^+\rangle\! \langle \text{GHZ}_N^+ | - |\text{GHZ}_N^-\rangle\! \langle \text{GHZ}_N^-|\right)\\
&\!\!\! \geq \!\!\!& \sqrt{2}^{N-1} \left[\sigma_x^{1} \hdots \sigma_x^N + \sum_{m\neq n}^N \sigma_z^{m}\sigma_z^n - N(N-1)\mathbbm{1} \right].
\label{bound_Mermin_Svetlichny}
\end{eqnarray}
It is not difficult to see that the same reasoning holds for odd $N$ (in this case, the basis is formed with all possible spin flips of $(N-1)/2$ spins), and consequently the above bound is valid for any $N$. 
Noticing then that $\sum_{m\neq n}^N \sigma_z^{m}\sigma_z^n=4S_z^2+N\mathbbm{1}$, where $S_z=(1/2)\sum_{i=1}^N\sigma_z^i$ is the total spin component along the $z$ axis, we arrive at the following operator bound for the Svetlichny and Mermin Bell operators
\begin{eqnarray}
\nonumber
\mathcal{B}_N^\text{Svet}=\mathcal{B}_N^\text{Mermin}&\!\!\!=\!\!\!&\sqrt{2}^{N-1}\left(|\text{GHZ}_N^+\rangle\! \langle \text{GHZ}_N^+ | - |\text{GHZ}_N^-\rangle\! \langle \text{GHZ}_N^-|\right)\\
&\!\!\! \geq \!\!\!& \sqrt{2}^{N-1} \left[\sigma_x^{1} \hdots \sigma_x^N + 4S_z^2 - N^2\mathbbm{1} \right].
\label{bound_Mermin_Svetlichny}
\end{eqnarray}
Combining the $k$-nonlocal bounds of the Svetlichny and Merming Bell expressions then allows us to write the following witness of Bell correlations depth:
\begin{equation}
\langle\mathcal{B}_N\rangle = \sqrt{2}^{N-1} \left\langle\sigma_x^{1} \hdots \sigma_x^N + 4S_z^2 - N^2\mathbbm{1} \right\rangle \leq 2^{(N-\lceil\frac{N}{k}\rceil)/2}.
\label{WitnessMS}
\end{equation}
Ineq. \eqref{WitnessMS} shows that two settings are enough to  conclude about the Bell correlation depth of a given state, that is, to test the capability of a state to violate a Svetlichny/Mermin bound for k-nonlocality. This provides a way to detect various depths of Bell correlations  with just two measurement settings and no individual addressing of the parties. In particular, since the GHZ state saturates all the inequalities we used in this section, the operator $\mathcal{B}_N$ is able to detect that GHZ states are genuinely Bell correlated.

Still, this scheme involves one parity measurement: the $N$-body term in the $x$ direction. It is worth noticing that the evaluation of this term does not require an estimation of all the moment of the spin operator $S_x$ in the $x$ direction (which would require a gigantic amount of statistics to be evaluated properly whenever $N\gg 1$). Rather, this term corresponds to the parity of the spin operator $S_x$, i.e. a binary quantity, and can thus be evaluated efficiently. However, an extreme resolution is required to estimate this quantity; failure to distinguish between two successive values of $S_x$ can entirely randomize its parity. 

The next section aims at detecting the nonlocality depth of multi-partite states with two-body correlators only.

\section{Characterising the sets of $k$-producible correlations with two-body correlators}
\label{Sec3}

Our aim in this section is the characterization of the two-body polytopes
of $k$-producible correlations defined in Sec.~\ref{k-producibility}. 
For this purpose, we will also determine
vertices of the projections of the nonsignaling polytopes onto two-body symmetric
correlations for a small number of parties.

\subsection{Two-body correlations}

Let us now imagine that we want to test nonlocality depth of some correlations, however, the only experimental data we have access to are the following one- and two-body expectation values
\begin{equation}\label{expvalues}
\langle M^{(i)}_k\rangle,\qquad \langle M^{(i)}_k M^{(j)}_l\rangle
\end{equation}
with $i\neq j=1,\ldots,N$ and $k,l=0,1$. As in the general case the most straightforward way of tackling this problem is to construct the corresponding polytope of $k$-producible correlations, denoted $\mathcal{P}_{N,k}^{2}$. This can be realized by projecting the polytope 
$\mathcal{P}_{N,k}$ onto the subspace of $\mathbbm{R}^{D}$ spanned by (\ref{expvalues}). In other words, the polytope $\mathcal{P}_{N,k}^2=\pi_2(\mathcal{P}_{N,k})$, where $\pi_2$ stands for the projection onto the two-body correlations, is obtained by getting rid of correlators (\ref{correlators}) of order larger than two in all elements of $\mathcal{P}_{N,k}$ (see also Appendix \ref{AppA} for a more detailed explanation of this projection). Having defined such a two-body polytope $\mathcal{P}_{N,k}^{2}$ for $k$-producible correlations we can address the problem of detecting nonlocality depth by constructing Bell-like inequalities constraining this polytope. However, to simplify this task, we can follow the approach of Ref. \cite{ScienceOur,JPA,Symmetric},
and consider inequalities that obey some symmetries, especially those that naturally arise in physical systems such as permutational or translational symmetry. Here we focus on the first one, that is we consider Bell-like expressions that are invariant under an exchange of any pair of parties. Mathematically, such a symmetrization is realized by another projection $\pi_{\mathrm{sym}}$ that maps one and two-body expectation values (\ref{expvalues}) onto the symmetric one and two-body correlations given by 
\begin{equation}\label{one}
\mathcal{S}_m := \sum_{i =1}^N \langle M_m^{(i)}\rangle,\qquad 
\mathcal{S}_{mn} := \sum_{\substack{i,j =1 \\ i \neq j}}^N \langle M_m^{(i)} M_n^{(j)}\rangle
\end{equation}
with $m,n=0,1$. 

By combining these two projections $\pi_2$ and $\pi_{\mathrm{sym}}$ we obtain another polytope which is our main object of study in this work: the two-body symmetric polytope of $k$-producible correlations $\mathcal{P}_{N,k}^{2,S}$. We want to provide inequalities that constrain $\mathcal{P}_{N,k}^{2,S}$ for various values of $k$, or, phrased differently, we want to find inequalities built from 
the symmetrized correlators (\ref{one}) that allow one to detect $k$-nonlocality in multipartite correlations. The most general form of such Bell-like inequalities is
\begin{equation}\label{k-inequalities}
I:=\alpha\mathcal{S}_0+\beta\mathcal{S}_1+\frac{\gamma}{2}\mathcal{S}_{00}+\delta \mathcal{S}_{01}+\frac{\varepsilon}{2} \mathcal{S}_{11}\geq -\beta_C^k,
\end{equation}
with 
\begin{equation}\label{kproduciblebounds}
\beta_C^k=-\min_{\mathcal{P}_{N,k}^{2,S}} I
\end{equation}
where the minimum is taken over all correlations belonging to $\mathcal{P}_{N,k}^{2,S}$; in fact, as the latter is a polytope it is enough to 
find the minimal value of $I$ over all its vertices in order to determine $\beta_C^k$. 
Notice that for our convenience in Eq. (\ref{k-inequalities}) we consider the lower $k$-nonlocal bound, whereas in Sec. \ref{Sec2bis} we considered bounds from above. For this reason, 
the bounds $\beta_{C}^k$ are defined in a different way than those in Eq. (\ref{BellLikeIn}). Nevertheless, we denote both bounds using the same letter because, conceptually, their meaning remains the same.

Note that $\mathcal{P}_{N,k}^{2,S}$ does not depend on the order in which the projections $\pi_2$ and $\pi_{\mathrm{sym}}$ are applied to $\mathcal{P}_{N,k}$. In other words, the following diagram is commutative:
\adjustbox{scale=1.5,center}{%
\begin{tikzcd}
{\cal P}_{N,k} \arrow[r, "\pi_2"] \arrow[d, "\pi_{\mathrm{sym}}"]
& {\cal P}_{N,k}^2 \arrow[d, "\pi_{\mathrm{sym}}"] \\
{\cal P}_{N,k}^S \arrow[r, "\pi_{2}"]
& {\cal P}_{N,k}^{2, S}
\end{tikzcd}
}
The result easily follows from the fact that each coordinate of $\mathcal{P}_{N,k}$ participates solely in one coordinate of $\mathcal{P}_{N,k}^{2,S}$ (cf. Eq. (\ref{one})) and $\pi_2, \pi_{\mathrm{sym}}$ are linear projections that act as follows: $\pi_2$ discards all correlators that involve more than $2$ parties and $\pi_{\mathrm{sym}}$ sums disjoint sets of correlators at each component.

Notice that for $k=1$, Eq. (\ref{k-inequalities}) reproduces the standard two-body symmetric Bell inequalities introduced and studied in \cite{ScienceOur}; in particular, in this case 
the vertices of $\mathcal{P}_{N,1}^{2,S}$ were fully characterized, which in turn made it possible to find a large part of the facets of $\mathcal{P}_{N,1}^{2,S}$, and thus two-body symmetric Bell inequalities detecting nonlocality in multipartite quantum states for any number of parties. Moreover, the algebraic structure present in the vertices of $\mathcal{P}_{N,1}^{2,S}$ has enabled the development of a hierarchy of SdP tests approximating this polytope from outside, which can be viewed as a method to check all Bell inequalities at once \cite{ThetaBodies}. 

Our aim in this work is to go beyond the case of $k=1$ and find inequalities valid for larger value of $k$ and any $N$. However, despite the fact that for any $k$ and $N$, the polytope $\mathcal{P}_{N,k}^{2,S}$ lives in a five-dimensional real space, its vertices and facets are unknown. Below we show how for a few values of $k$, the vertices of the symmetric two-body polytope can be determined.

\subsection{Characterization of the vertices of the $k$-nonlocal two-body symmetric polytopes}\label{sec:kNvertices}

Here we introduce a general description of all the vertices of the projected $k$-nonlocal polytope in the two-body symmetric space. This description is in principle valid for any value of $N$ and $k \leq N$, however, requires knowing all the vertices of the symmetrized local polytope for $p$ parties with $p=1,\ldots,k$ and all the nonlocal vertices of the projections of $p$-partite nonsignaling polytopes $\mathcal{NS}_p$ onto the two-body symmetric space for $p=1,\ldots,k$, which we denote $\mathcal{NS}_{p}^{2,S}$. In the next subsection we show how to find their vertices for a few values of $k$.

Let us use the following notation. By $\vec{S}(p,i)$
we denote a five-dimensional vector 
\begin{equation}\label{5dimvector}
\vec{S}(p,i)=(S_0(p,i),S_{1}(p,i),S_{00}(p,i),S_{01}(p,i),S_{11}(p,i))
\end{equation}
of one- and two-body symmetric expectation values for
the $i$th vertex of $\mathcal{NS}_{p}^{2,S}$ with $i=1,\ldots,n_p$
and $p=1,\ldots,k$, where $n_p$ stands for the number of vertices of $\mathcal{NS}_{p}^{2,S}$. 
Let then $\{\vec{S}(p,i)\}_{p,i}$ be the list of all such five-dimensional vectors (\ref{5dimvector}).

Then, each vertex of the two-body symmetric polytope of $k$-producible correlations $\mathcal{P}_{N,k}^{2,S}$, which to recall is a projection of a vertex (\ref{vertex}) onto the two-body symmetric subspace, can be parametrized by the populations $\xi_{p,i}$ with  
$p = 1,\ldots,k$, representing the number of $p$-partite subgroups $\mathcal{A}_l$ in the $k$-partition of the set $\{A_1,\ldots,A_N\}$ (cf. Sec. \ref{k-producibility}) that are adopting the ``strategy" from the list $\{\vec{S}(p,i)\}_{p,i}$. Indeed, since we are addressing permutationally invariant quantities, we are insensitive to the assignment of a strategy to a specific group of parties, hence the only relevant information is about the number of parties adopting each given set of correlations.

\begin{figure}[t]
\centering
\includegraphics[scale=.85]{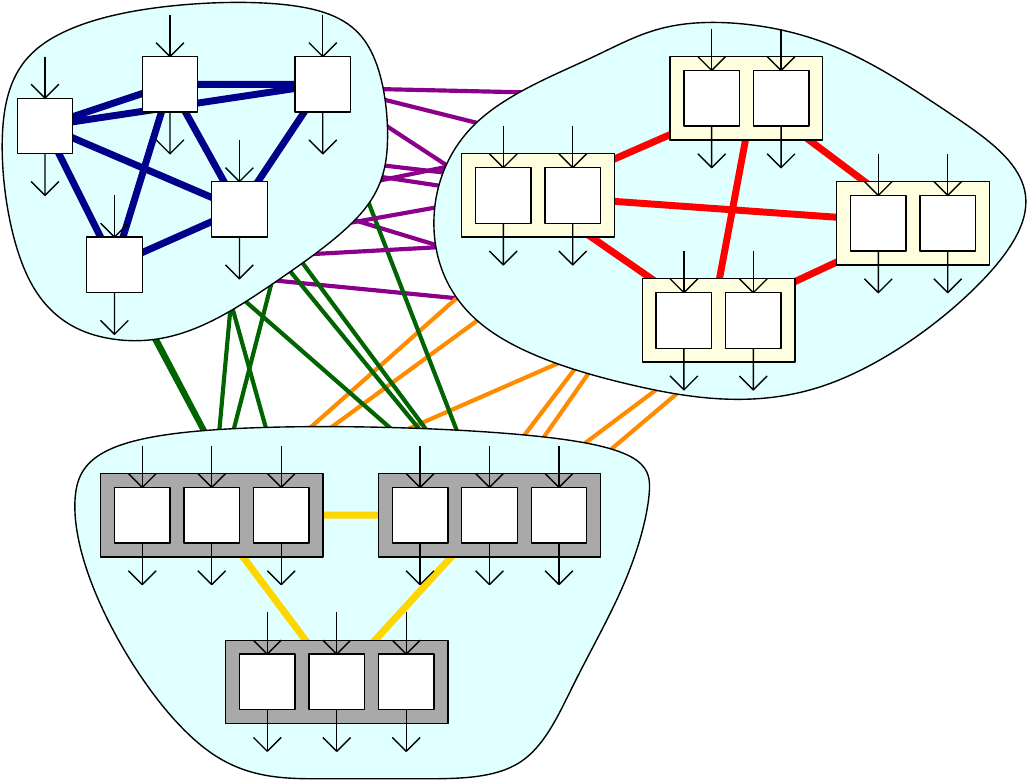}
\caption{An example with $N=22$, and the $3$-partition into $n_1 = 5$ sets of size $1$, $n_2 = 4$ of size $2$ and $n_3 = 3$ of size $3$. The first sum in Eq. (\ref{Smn}) corresponds to the value of $S_{mn}$ that comes from the two-body correlators $\langle M_{m}^{(i)}M_n^{(j)}\rangle$ within each set (i.e., $i,j\in {\cal A}_l$ for some $l$). For the $1$-body boxes, these values are clearly zero, and for larger boxes, they correspond to the two-body marginals of the corresponding Popescu-Rohrlich box (PR-box). Therefore, once symmetrized, the contribution of the box involving $p$ parties using the $i$-th strategy is $S_{mn}(p,i)$. The second sum in Eq. (\ref{Smn}) counts those two-body correlators $\langle M_{m}^{(i)}M_n^{(j)}\rangle$ in which $i\in {\cal A}_k$, $j\in {\cal A}_l$, $k \neq l$ and $|{\cal A}_k| = |{\cal A}_l| = p$. These correlations are represented by the blue, red and yellow lines. Because they are correlations coming from different PR-boxes, the locality assumption guarantees the factorization $\langle M_{m}^{(i)}M_n^{(j)}\rangle=\langle M_m^{(i)}\rangle \langle M_n^{(j)}\rangle$, yielding the term $S_m(p,i)S_n(p,i)$ once symmetrized. The factor $\xi_{p,i}(\xi_{p,i}-1)$ is given by the fact that ${\cal S}_{mn}$ is defined as the sum for all $i \neq j$, therefore containing repetitions. Finally, the last sum in Eq. (\ref{Smn}) is given by all two-body correlators $\langle M_{m}^{(i)}M_n^{(j)}\rangle$ in which $i\in {\cal A}_k$, $j\in {\cal A}_l$, and $|{\cal A}_k| = p$, $|{\cal A}_l| = q$ with $p \neq q$, i.e., two-body correlations connecting PR-boxes of different size. Here the locality assumption also enables a factorization which amounts to $S_m(p,i)S_n(q,j)$ once symmetrized, weighted by the number of ocurrences $\xi_{p,i}\xi_{q,j}$. These correspond to the purple, green and orange lines in the figure.}
\label{fig.groups}
\end{figure}

Now, as parties are divided into subsets of size at most $k$, and each subset of size $p$ chooses one out of $n_p$ strategies, 
the populations $\xi_{p,i}$, weighthed by $p$, form a partition of $N$, that is,
\begin{equation}\label{condition}
\sum_{p=1}^k \sum_{i=1}^{n_p}p\,\xi_{p,i}=N.
\end{equation}
By running over all populations $\xi_{p,i}$ obeying (\ref{condition}), one spans the whole set of vertices of the polytope $\mathcal{P}_{N,k}^{2,S}$. Moreover, denoting by $\vec{\xi}$ the vector of populations $\xi_{p,i}$,
the symmetrized one- and two-body expectation values for the vertices 
of $\mathcal{P}_{N,k}^{2,S}$ can be expressed as
\begin{equation}\label{Sm}
S_m(\vec{\xi})=\sum_{p=1}^{k}\sum_{i=1}^{n_p}\xi_{p,i}S_m(p,i)
\end{equation}
and
\begin{eqnarray}\label{Smn}
S_{mn}(\vec{\xi})&\!\!\!=\!\!\!&\sum_{p=1}^{k}\sum_{i=1}^{n_p}\xi_{p,i}S_{mn}(p,i)+
\sum_{p=1}^k\sum_{i=1}^{n_p}\xi_{p,i}(\xi_{p,i}-1)S_m(p,i)S_n(p,i)\nonumber\\
&&+\sum_{\{p,i\}\neq \{q,j\}}\xi_{p,i}\xi_{q,j}S_m(p,i)S_n(q,j),
\end{eqnarray}
where we used the fact that $\langle M_{m}^{(i)}M_n^{(j)}\rangle=\langle M_m^{(i)}\rangle \langle M_n^{(j)}\rangle$ whenever the parties $i$ and $j$ belong to different groups, and
$\{p,i\}\neq \{q,j\}$ means that $p\neq q$ or $i\neq j$. Fig. \ref{fig.groups} 
explains how the different correlators are summed in formula (\ref{Smn}).
%

\subsection{Projecting the nonsignaling polytopes}

As already said, in order to generate the vertices of the symmetric two-body polytope of $k$-producible correlations $\mathcal{P}_{N,k}^{2,S}$, we need to know the vertices of the nonsignaling polytope $\mathcal{NS}_p$ in the two-body symmetric space for $2\leq p \leq k$ parties. To this aim, we need to determine its projection $\mathcal{NS}^{2,S}_p$ onto the two-body symmetric space spanned by (\ref{one}) for any $p=2,\ldots,k$. Below we determine all these vertices for $k=2,\ldots,6$.

\subsubsection{Projecting through the vertices. The $k = 2,3$ cases}

For the smallest $k$'s, the lists of vertices of the nonsignalling polytopes are known. 
For $k = 2$, the only nonlocal vertices belong to the equivalence class of the so-called PR-box \cite{PRbox}. For $k = 3$, the list of the $46$ equivalence classes was derived in \cite{NS}.
Therefore, for these scenarios, the projection can be performed straightforwardly through the vertex representation (cf. Appendix \ref{AppA}). The resulting extremal points are shown in Table \ref{table2} and \ref{table3} in Appendix \ref{app:vertices} (the vertices that are shared with the local polytope are omitted).

For the bipartite case, $k=2$, the $4$ nontrivial vertices belong obviously to a single equivalence class, corresponding to the projection of the PR-box.
Interestinlgy, there is also only one relevant class for the tripartite case, corresponding to the projection of the class number $29$ of \cite{NS}, which is one of the two that violate maximally the Guess-Your-Neighbour-Input inequality \cite{GYNI}.

\subsubsection{A method to find the projection for $k \geq 4$}

The vertices of $\mathcal{NS}_k$ for $k>3$ are unknown and difficult to determine. However, as already mentioned, its facets are easy to describe by the inequalities (\ref{posit}), which in the correlators picture can be stated as [cf. Eq. (\ref{relation})]:
\begin{equation}
 \sum_{k = 1}^N \sum_{1 \leq i_1<\ldots<i_k \leq N} a_{i_1}\ldots a_{i_k} \big\langle M^{(i_1)}_{x_{i_1}} \ldots M^{(i_k)}_{x_{i_k}} \big\rangle+1 \geq 0,
\label{pos}
\end{equation}
for all the possible outcomes $a_{i_1},\ldots,a_{i_N} = \pm 1$ and measurement choices $x_{1},\ldots,x_{N} = 0,1$. 
(Recall that the normalization and the no-signaling conditions are already included in this picture.)

Thus, an approach to find the projections
of $\mathcal{NS}_k$ for $k>3$ would be to use the Fourier-Motzkin procedure (see Appendix \ref{AppA}). 
However, given that the dimension $D(k)=3^k-1$ of the space of no-signalling correlations scales exponentially with the number of parties $k$, the projection $\pi_2\circ \pi_{\mathrm{sym}}: \mathbb{R}^{D(k)} \rightarrow \mathbb{R}^{5} $ \textit{via} the Fourier-Motzkin method requires eliminating an exponentially increasing number of variables. Due to this limitation, the general approach of projecting the no-signalling polytope into the two-body symmetric space becomes impractical already for the case $k = 3$. In what follows we will show how to overcome this difficulty and find the 
vertices for $k=4,5,6$. 

To this end, let us denote by $V_2$ the subspace of $\mathbbm{R}^{D(k)}$ 
spanned by one- and two-body expectation values (\ref{expvalues}) and by 
$V_{\mathrm{sym}}$ the subspace of $\mathbbm{R}^{D(k)}$ spanned by the symmetrized
correlators of any order:
\begin{equation}
S_{m_1\ldots m_l}=\sum_{i_1\neq \ldots \neq i_l=1}^{k}\big\langle M_{m_1}^{(i_1)}\ldots M_{m_l}^{(i_l)}\big\rangle
\end{equation}
with $m_i=0,1$ and $l=1,\ldots,k$. We also introduce the coordinates
\begin{equation}
T_{m_1\ldots m_l}^{j_1\ldots j_l} = \left(\sum_{i_1\neq \ldots \neq i_l=1}^{k}\right) \langle M_{m_1}^{(j_1)}\ldots M_{m_l}^{(j_l)}\big\rangle - S_{m_1\ldots m_l}.
\end{equation}
Taken together, the $S$ and $T$ correlators provide an over-complete parametrization of the no-signaling probability space: any correlator $\langle M_{m_1}^{(j_1)}\ldots M_{m_l}^{(j_l)}\big\rangle$ can be recovered from the corresponding $T$ and $S$ variables. Moreover, these coordinates conveniently identify the subspaces that we are interested in: the symmetric subspace is spanned by the $S$ variables, while its orthogonal complement by the $T$ variables. 
In other words, all correlations in the symmetric subspace have all $T$ components equal to zero (however, the variables themselves before projection need not be zero). Moreover, the one- and two-body space has all $S$ and $T$ parameters equal to 0 for $l>2$.

We then notice that the projection we want to compute can be divided into two intermediate steps as $P = \pi_{2} \circ \pi_{\mathrm{sym}}  =  \pi_{\mathrm{sym}} \circ \pi_{2} $, where $\pi_{2}$ and $\pi_{\mathrm{sym}}$ stand for projections onto $V_2$ and $V_{\mathrm{sym}}$, respectively.

At this stage, it is worth noting that the set that we wish to project, $\mathcal{NS}_N$ is invariant under parties permutation. This implies that the projection of this set onto $V_{\mathrm{sym}}$ coincides with the intersection between $\mathcal{NS}_N$ and $V_{\mathrm{sym}}$, i.e.
\begin{equation}
\mathrm{int}_{\mathrm{sym}}(\mathcal{NS}_k) =  \pi_{\mathrm{sym}}(\mathcal{NS}_k),
\end{equation}
where, within the above parametrization, the set $\mathrm{int}_{\mathrm{sym}}(\mathcal{NS}_k)$ contains those correlations for which all variables $T_{m_1\ldots m_l}^{j_1\ldots j_l}=0$.
Indeed, if a point $p$ in $\mathcal{NS}_N$ (which may have both non-zero $S$ and $T$ components) leads to an extremal vertex after projection onto $V_{\mathrm{sym}}$, then all images $p_\alpha=\tau_\alpha(p)$ of $p$ under the party permutations $\{\tau_\alpha\}_\alpha$ are also in $\mathcal{NS}_N$, and lead to the same point in $V_{\mathrm{sym}}$ after projection. Therefore, the convex combination of these points $\overline p \propto \sum_\alpha p_\alpha$ also gives rise to the same extremal point in $V_{\mathrm{sym}}$. However, a direct computation shows that the point $\overline p$ already belongs to the symmetric subspace, because all of its $T$ variables  are zero. Hence, all extremal points of the projection of $\mathcal{NS}_N$ onto the symmetric subspace belong to the intersection of the no-signaling polytope with the symmetric subspace, and we can replace the projection operation $\pi_{\mathrm{sym}}$ by the intersection.

The main advantage of this approach is that we now only need to apply the Fourier-Motzkin method to perform the projection onto the two-body space $V_2$. In this case  the number of variables to discard does not grow exponentially with $N$. Indeed, the number of symmetric correlators $S_{m_1...m_l}$ with $m_j=0,1$ and $l=1,\ldots,k$ scales as $(1/2)(N+1)(N+2)-1$ and, since one has to discard all the terms with $k>2$, we need eliminate $(1/2)(N+1)(N+2)-6  \approx \mathcal{O}(N^2)$ terms.

For the $k=2,3,4$ cases the lists of vertices are presented in Tables \ref{table2}--\ref{table4} in Appendix \ref{app:vertices}, whereas in the case $k=5,6$ the list contains more than a hundred vertices and therefore we could not present it here. In the section that follows we implement these findings to construct Bell-like inequalities detecting $k$-nonlocality in multipartite correlations.

\section{Bell-like inequalities for nonlocality depth from two-body correlations}
\label{Sec4}

We are now ready to demonstrate that two-body Bell-like inequalities are capable of witnessing nonlocality depth in multipartite correlations.
First of all, we remind that, by following the procedure given in section \ref{sec:kNvertices}, we are able to construct the list of vertices of the $k$-nonlocal two-body symmetric polytopes for any number of parties $N$ and producibility $k \leq 6$. 
By solving the convex hull problem, such lists allow us to derive the corresponding complete set of facets of the $k$-nonlocal polytopes. This can be done via dual description method, which is implemented in such software as CDD \cite{Fukuda}, and, thanks to the low dimension of the space, we are able to do so for scenarios involving up to $N = 12$ parties.

In particular, since these inequalities can test against $k$-producibility with $k \leq 6$, we can identify all the symmetric two-body inequalities that detect genuine multipartite nonlocality (GMNL) for systems of $N \leq 7$ particles (see Appendix \ref{app:facets} for the complete lists).  
Interestingly, we find that no inequality of such kind can be violated by quantum mechanics in the tripartite case. That is, symmetric two-body correlations provide not enough information to detect GMNL in three-partite quantum states. This is no longer the case for four parties; indeed, the following facet Bell-like inequality
\begin{equation}
-12 S_0 + 9 S_1 + 3 S_{00} - 6 S_{01} + \frac{1}{2} S_{11} + 42 \geq 0
\end{equation}
detects GMNL and is violated by quantum mechanics with a ratio $(\beta_Q - \beta_C^3)/\beta_C^3$ of at least $1.3 \%$, where $\beta_Q$ is the maximal quantum value of the corresponding Bell expression.

Interestingly, our lists of inequalities sometimes contain also the Bell expressions introduced already in \cite{ScienceOur,Series2}, thus showing that such classes are actually capable of detecting a nonlocality depth higher than two. In particular, we can find inequalities that test against any $k$-producibility for $k \leq 5$ that belong to the class (91) introduced in Ref. \cite{Series2}. This class is particulary interesting since it was shown to be violated by Dicke states. Moreover, among the facets of the GMNL polytope for $N = 5$, we find the following inequality 
\begin{equation}
28 S_0 + 28 S_1 + 2 S_{00} + 9 S_{01} + 2 S_{11} + 116 \geq 0
\end{equation}
which has a very similar structure to class (91) of Ref. \cite{Series2}. Indeed, it can be shown that it is possible to violate such inequality with the five-partite Dicke state with one excitation, also known as the $W$ state.
Lastly, we notice that the Bell expression (6) from Ref. \cite{ScienceOur}, which for the sake of completeness we state here as,
\begin{equation}\label{expression6}
\mathcal{I}:=2S_0+\frac{1}{2}S_{00}+S_{01}+\frac{1}{2}S_{11}
\end{equation}
appears in our lists sometimes as well, with a classical bound that clearly depends on degree of nonlocality depth that one is interested to detect. This is a particularly useful feature, since it implies that by the use of a single inequality one can infer the nonlocality depth by the amount of the quantum violation that is observed. Due to this property and also its relevance for experimental implementation (see Refs. \cite{Schmied,Engelsen}), in the next section we focus on this last inequality and determine its $\beta_{C}^k$ for $k=2,\ldots,6$ and any number of parties.

\section{Deriving the inequalities and witnesses of $k$-nonlocality for any number of parties}
\label{Sec5}

We are now ready to show that one can reveal $k$-nonlocality for some values of $k$
or test the nonlocality depth only from the two-body correlations for any number of parties.
To that end, in what follows we compute 
$k$-producible bounds $\beta_C^{k}$ [cf. Eq. (\ref{kproduciblebounds})]
for different $k$'s for the two-body Bell expression (\ref{expression6}).
We begin with fully general considerations and later we focus on a few values of $k$ and compute $\beta_C^k$ case by case. Due to the fact that $\mathcal{P}_{N,k}^{2,S}$ is a polytope, it is enough to perform the above minimization over its vertices. Using Eqs. (\ref{Sm}) and (\ref{Smn}), the expression $\mathcal{I}$ in Eq. (\ref{expression6}) 
for all vertices of $\mathcal{P}_{N,k}^{2,S}$ can be written as
\begin{align}
\mathcal{I}(\vec{S}( \vec{\xi})) = & \sum_{p = 1}^k \sum_{i=1}^{n_p} \xi_{p,i} \mathcal{I} (\vec{S}(p,i)) +  \frac{1}{2}  \sum_{p = 1}^k \sum_{i=1}^{n_p} \xi_{p,i} \left[\xi_{p,i} - 1\right] \mathcal{I} (\vec{S}(p,i) ,\vec{S}(p,i) ) + \nonumber \\
& + \frac{1}{2}  \sum_{ \lbrace p,i \rbrace \neq  \lbrace q,j \rbrace} \xi_{p,i} \xi_{q,j} \mathcal{I} (\vec{S}(p,i) ,\vec{S}(q,j) ),
\label{poly}
\end{align}
where we have defined the following cross-terms
\begin{equation}
\mathcal{I}(\vec{S}(p,i),\vec{S}(q,j)) = \left[S_0(p,i) + S_1 (p,i)\right]\left[S_0(q,j) + S_1 (q,j)\right].
\label{cross}
\end{equation}

When the vectors $\vec{S}(p,i)$ are known, the expression (\ref{poly}) takes the form of a polynomial of degree two in terms of the populations $\xi_{p,i}$. By grouping together the linear and quadratic terms, we get
\begin{align}
\mathcal{I} (\vec{S}( \vec{\xi})) = & \sum_{p =1}^k \sum_{i=1}^{n_p} \xi_{p,i} \left[ \mathcal{I} (\vec{S}(p,i))  - \frac{1}{2} \mathcal{I} (\vec{S}(p,i) ,\vec{S}(p,i) ) \right] + \nonumber \\
& \frac{1}{2} \sum_{p,q=1}^k \sum_{i,j=1}^{n_p} \xi_{p,i} \xi_{q,j} \mathcal{I} (\vec{S}(p,i) ,\vec{S}(q,j) ) 
\end{align}
Then, by substituting the explicit form of the cross-term (\ref{cross}), one arrives at
\begin{align}
\mathcal{I} (\vec{S}( \vec{\xi})) = & \sum_{p = 1}^k \sum_{i=1}^{n_p} \xi_{p,i} \left\{  \mathcal{I} (\vec{S}(p,i))  - \frac{1}{2} [S_0 (p,i) + S_1(p,i)]^2 \right\} + \nonumber \\
& \frac{1}{2} \left\{ \sum_{p = 1}^k \sum_{i=1}^{n_p} \xi_{p,i} [S_0 (p,i) + S_1 (p,i)] \right\}^2 .
\label{poly2}
\end{align}
With the above expression at hand we can now seek the $k$-producibility bounds $\beta_C^k$ for $\mathcal{I}$. Our approach is the following. Instead of minimizing the expression $\mathcal{I}$ for all $k$-producible correlations, we will rather consider a particular value of $\beta_C^k$ and prove that the inequality $\mathcal{I}+\beta_C^k\geq 0$ holds for all integer values of $\xi_{p,i}\geq 0$ for $p=1,\ldots,k$ and $i=1,\ldots,n_p$ such that the condition (\ref{condition}) holds.

\subsection{Cases $k=2$ and $k=3$}

We will first consider the simplest cases of $k=2,3$ and show that for them $\beta_{C}^k=2N$ is the correct classical bound. In other words, below we demonstrate the the following 
inequality
\begin{equation}
2S_0+\frac{1}{2}S_{00}+S_{01}+\frac{1}{2}S_{11}+2N\geq 0
\label{eq:ineq6}
\end{equation}
is satisfied for all correlations belonging to $\mathcal{P}_{N,k}^{2,S}$ for $k=2,3$ and any $N$. 
To this end, we use Eqs. (\ref{poly2}) and (\ref{condition}) to write down the following expression 
\begin{eqnarray}\label{TintoDeVerano}
\mathcal{I} (\vec{S}( \vec{\xi})) +2N&\!\!\!=\!\!\!& \sum_{p = 1}^k \sum_{i=1}^{n_p} \xi_{p,i} \left\{  \mathcal{I} (\vec{S}(p,i)) +2p - \frac{1}{2} [S_0 (p,i) + S_1(p,i)]^2 \right\} + \nonumber \\
&& \frac{1}{2} \left\{ \sum_{p = 1}^k \sum_{i=1}^{n_p} \xi_{p,i} [S_0 (p,i) + S_1 (p,i)] \right\}^2 .
\end{eqnarray}
Then, plugging in the explicit values of the one- and two-body symmetric expectation values for $p=1,2,3$ collected in Tables \ref{table1}--\ref{table3}, the above further rewrites as
\begin{eqnarray}\label{expression112}
\mathcal{I} (\vec{S}( \vec{\xi})) +2N&\!\!\!=\!\!\!&2\left[\left(\xi_{1,1}-\xi_{1,4}-\xi_{3,1}-\xi_{3,2}+\xi_{3,7}+\xi_{3,8}\right)^2+\xi_{1,1}+2\xi_{1,2}-\xi_{1,4}\right]\nonumber\\
&&+2\left[3\left(\xi_{2,1}+\xi_{2,2}\right)+\xi_{2,3}+\xi_{2,4}\right]\nonumber\\
&&+2\left[\left(\xi_{3,1}+\xi_{3,2}\right)+4\left(\xi_{3,3}+\xi_{3,4}\right)+6\left(\xi_{3,5}+\xi_{3,6}\right)+3\left(\xi_{3,7}+\xi_{3,8}\right)\right].
\end{eqnarray}
With the following substitutions 
\begin{eqnarray}
\mathcal{X}&\!\!\!=\!\!\!&2\left(\xi_{1,1}-\xi_{1,4}\right),\nonumber\\
\mathcal{Y}&\!\!\!=\!\!\!&2\left(-\xi_{3,1}-\xi_{3,2}+\xi_{3,7}+\xi_{3,8}\right),\nonumber\\
\mathcal{P}(\vec{\xi})&\!\!\!=\!\!\!&2\left[\xi_{1,2}+3\left(\xi_{2,1}+\xi_{2,2}\right)+\xi_{2,3}+\xi_{2,4}+2\left(\xi_{3,1}+\xi_{3,2}\right)\right.\nonumber\\
&&\left.+4\left(\xi_{3,3}+\xi_{3,4}\right)+6\left(\xi_{3,5}+\xi_{3,6}\right)+2\left(\xi_{3,7}+\xi_{3,8}\right)\right],
\end{eqnarray}
we can bring the expression (\ref{expression112}) into the following simple form
\begin{eqnarray}\label{NJP}
\mathcal{I} (\vec{S}( \vec{\xi})) +2N&\!\!\!=\!\!\!&\frac{1}{2}\left(\mathcal{X}+\mathcal{Y}\right)^2+\mathcal{X}+\mathcal{Y}+\mathcal{P}(\vec{\xi})\nonumber\\
&\!\!\!=\!\!\!&2\mathcal{Z}(\mathcal{Z}+1)+\mathcal{P}(\vec{\xi}),
\end{eqnarray}
where $\mathcal{Z}=(\mathcal{X}+\mathcal{Y}/2)$. We then notice that all $\xi_{p,i}\geq 0$, which immediately implies that $\mathcal{P}(\vec{\xi})\geq 0$ for all configurations of populations. Thus, we are left with the term $\mathcal{Z}(\mathcal{Z}+1)$, which is negative only when $-1<\mathcal{Z}<0$. However, due to the fact that $\mathcal{Z}$ is a linear combination of integers with integer coefficients, it cannot take such values. Thus, 
$\mathcal{Z}(\mathcal{Z}+1)\geq 0$, which completes the proof.

\subsection{The case $k=4$}

We now address the first case in which the bound $\beta_C^k$ is different than the local bound of the Bell inequality. First of all, we notice that the bipartite nonsignaling populations $\xi_{2,i}$ enter the expression (\ref{TintoDeVerano}) only in the linear part and, since their coefficient are always positive, we know that they never contribute to the violation of the local bound. Thus, the expression $\mathcal{I}(\vec{S}(\vec{x}))+2N$ without these terms reads explicitly
\begin{eqnarray}
&&\hspace{-1cm}\frac{1}{2}\left(\mathcal{X}+\mathcal{Y}+\mathcal{Y}'+\mathcal{W}+\mathcal{W}'\right)^2+2\left(\xi_{1,1}+2\xi_{1,2}-\xi_{1,4}\right)\nonumber\\
&&+2\left[\left(\xi_{3,1}+\xi_{3,2}\right)+4\left(\xi_{3,3}+\xi_{3,4}\right)+6\left(\xi_{3,5}+\xi_{3,6}\right)+3\left(\xi_{3,7}+\xi_{3,8}\right)\right]\nonumber\\
&&+2\left[\xi_{4,1}+\xi_{4,2}+5\left(\xi_{4,3}+\xi_{4,4}\right)+9\left(\xi_{4,5}+\xi_{4,6}\right)+5\left(\xi_{4,7}+\xi_{4,8}\right)\right]\nonumber\\
&&+8\left[\xi_{4,9}+\xi_{4,10}+2\left(\xi_{4,11}+\xi_{4,12}\right)\right]\nonumber\\
&&+\frac{8}{49}\left[-22\left(\xi_{4,13}+\xi_{4,14}\right)+19\left(\xi_{4,15}+\xi_{4,16}\right)+89\left(\xi_{4,17}+\xi_{4,18}\right)+48\left(\xi_{4,19}+\xi_{4,20}\right)\right],
\end{eqnarray}
where $\mathcal{X}$ and $\mathcal{Y}$ are defined above and $\mathcal{Y}'$, $\mathcal{W}$ and $\mathcal{W}'$ are given by
\begin{eqnarray}
\mathcal{Y'}&\!\!\!=\!\!\!&4\left(-\xi_{4,1}-\xi_{4,2}+\xi_{4,7}+\xi_{4,8}\right),\nonumber\\
\mathcal{W}&\!\!\!=\!\!\!&\frac{24}{7}\left(-\xi_{4,13}-\xi_{4,14}+\xi_{4,19}+\xi_{4,20}\right),\nonumber\\
\mathcal{W}'&\!\!\!=\!\!\!&\frac{16}{7}\left(-\xi_{4,15}-\xi_{4,16}+\xi_{4,17}+\xi_{4,18 }\right).
\end{eqnarray}
Then, we can simplify this expression 
\begin{equation}
\frac{1}{2}\left(\mathcal{X}+\mathcal{Y}+\mathcal{Y}'+\mathcal{W}+\mathcal{W}'\right)^2+\mathcal{X}+\mathcal{Y}+\mathcal{Y}'+\mathcal{W}+\mathcal{W}'+\widetilde{\mathcal{P}}(\vec{\xi})-\frac{8}{49}\left(\xi_{4,13}+\xi_{4,14}\right),
\end{equation}
where 
\begin{eqnarray}\label{CFT}
\widetilde{\mathcal{P}}(\vec{\xi})&\!\!\!=\!\!\!&4\xi_{2,2}+4\left[\xi_{3,1}+\xi_{3,2}+2\left(\xi_{3,3}+\xi_{3,4}\right)+3\left(\xi_{3,5}+\xi_{3,6}\right)+\xi_{3,7}+\xi_{3,8}\right]\nonumber\\
&&+2\left[3\left(\xi_{4,1}+\xi_{4,2}\right)+5\left(\xi_{4,3}+\xi_{4,4}\right)+9\left(\xi_{4,5}+\xi_{4,6}\right)+3\left(\xi_{4,7}+\xi_{4,8}\right)\right]\nonumber\\
&&+8\left[\xi_{4,9}+\xi_{4,10}+2\left(\xi_{4,11}+\xi_{4,12}\right)\right]\nonumber\\
&&+\frac{8}{49}\left[33\left(\xi_{4,15}+\xi_{4,16}\right)+75\left(\xi_{4,17}+\xi_{4,18}\right)+27\left(\xi_{4,19}+\xi_{4,20}\right)\right]
\end{eqnarray}
is a polynomial that is positive for all configurations of populations $\xi_{p,i}$. 
%
%
Let us now show that the expression in (\ref{CFT}) is always greater or equal to $-2N/49-1/2$. In other words, we want to prove that 
\begin{equation}\label{CFT2}
\frac{1}{2}\left(\mathcal{X}+\mathcal{Y}+\mathcal{Y}'+\mathcal{W}+\mathcal{W}'\right)^2+\mathcal{X}+\mathcal{Y}+\mathcal{Y}'+\mathcal{W}+\mathcal{W}'+\widetilde{\mathcal{P}}(\vec{\xi})-\frac{8}{49}\left(\xi_{4,13}+\xi_{4,14}\right)+\frac{1}{2}+\frac{2}{49}N\geq 0,
\end{equation}
for any $\xi_{p,i}$. To this end, we can exploit (\ref{condition}) in order to express $N$ in terms of the populations, which allows us to see that  $2N \geq 8(\xi_{4,13}+\xi_{4,14})$, implying that (\ref{CFT2}) holds true. As a result, 
the bound for $k=4$ amounts to 
\begin{equation}
\beta_C^4=\left(2+\frac{2}{49}\right)N+\frac{1}{2}.
\end{equation}

\subsection{Cases $k=5,6$}

Based on the previous results our guess is that for any $3<k<N$, the bound for $k$-producible correlations is given by
\begin{equation}\label{boundCk}
\beta_C^{k}=2N+\frac{1}{2}+\alpha_kN.
\end{equation}
In what follows we estimate the correction to the linear dependence on $N$, that is, $\alpha_k$ for $k=5,6$, and leave the general case of any $k$ as an open problem. To this aim, we follow the approach used in the case $k=4$. More precisely, by substituting $\beta_{C}^{k}$ given in (\ref{boundCk}) into $\mathcal{I}+\beta_C^{k}$, we obtain
\begin{eqnarray}
\mathcal{I}(\vec{S}( \vec{\xi}))+\beta_C^{k}&\!\!\!=\!\!\!&  \sum_{p = 1}^k \sum_{i=1}^{n_p} \xi_{p,i} \left[  \mathcal{I} (\vec{S}(p,i)) + \left(2 + \alpha_k \right) p  - \frac{1}{2} \left[S_0 (p,i) + S_1(p,i)\right]^2   \right]  \nonumber \\
&&+ \frac{1}{2} \left\{ \sum_{p = 1}^k \sum_{i=1}^{n_p} \xi_{p,i} \left[S_0 (p,i) + S_1 (p,i)\right] \right\}^2 + \frac{1}{2} \nonumber \\
&\!\!\!=\!\!\!&  \sum_{p = 1}^k \sum_{i=1}^{n_p} \xi_{p,i} \left\{  \mathcal{I}_6 (\vec{S}(p,i)) + \left(2 + \alpha_k \right) p  \right. \nonumber\\
&& \left.\hspace{2.2cm}- \frac{1}{2} \left[S_0 (p,i) + S_1(p,i)\right]^2-\left[S_0 (p,i) + S_1(p,i)\right] \right\}  \nonumber \\
 &&+ \frac{1}{2} \left\{ \sum_{p = 1}^k \sum_{i=1}^{n_p} \xi_{p,i} \left[S_0 (p,i) + S_1 (p,i)\right] + 1 \right\}^2. 
\label{poly4}
\end{eqnarray} 
As the last term in this expression is always nonnegative, to study the positivity of $\mathcal{I}(\vec{S}(\vec{\xi}))+\beta_C^{(k)}$, we can restrict our attention to 
the following function
\begin{eqnarray}
\Omega(\vec{\xi})&\!\!\!=\!\!\!&\sum_{p = 1}^k \sum_{i=1}^{n_p} \xi_{p,i} \left\{  \mathcal{I} (\vec{S}(p,i)) + \left(2 + \alpha_k \right) p  \right. 
 \left.- \frac{1}{2} \left[S_0 (p,i) + S_1(p,i)\right]^2-\left[S_0 (p,i) + S_1(p,i)\right] \right\}.\nonumber\\
\end{eqnarray}
As it is a linear function in the populations which are all nonnegative, its minimum is reached for the population $\xi^*$ standing in front the expression that takes the minimal value over all choices of the vertices. In other words, 
we can lower bound $\Omega(\vec{x})$ as
\begin{equation}
\Omega(\vec{\xi})\geq   \xi^*(p^*\alpha_k+m_k),
\end{equation}
where $p^*$ is the number of parties corresponding to $x^*$ and $m_k$ is defined as
\begin{equation}
m_k = \displaystyle{\min_{p=1,\ldots,k} \min_{i=1,\ldots, n_p} } \left\{  \mathcal{I} (\vec{S}(p,i)) + 2p  - \frac{1}{2} [S_0 (p,i) + S_1(p,i)]^2  -  [S_0 (p,i) + S_1(p,i)] \right\}.
\label{min}
\end{equation}
Thus, we simply need to compute $m_k$ and the value $\alpha_k$ we are looking for can be taken as $\alpha_k=m_k/k$ as for it $\Omega(\vec{\xi})\geq 0$ for any $N$.

To this end, we perform the minimization in 
(\ref{min}) by evaluating the right-hand side on each vertex of the projected five- and six-partite no-signaling polytope. We obtain $m_5=40/121$ and $m_6 = 1/2$, implying that the modified bounds amount to 
\begin{equation}
\beta_{C}^{5}=\left(2+\frac{8}{121}\right)N+\frac{1}{2},
\end{equation}
and
\begin{equation}
\beta_{C}^{6}=\left(2+\frac{1}{12}\right)N+\frac{1}{2},
\label{eq:beta6-conjecture}
\end{equation}
respectively.
\subsection{Quantum violations}

After having derived $k$-producible bounds for inequality (\ref{expression6}), we proceed to show that they can be used in pratice to witness the nonlocality depth that could be displayed by quantum states.
First of all, we have to show that the different bounds $\beta_C^{k}$ can be violated by correlations obtained by properly choosing a quantum state and some local measurements. This can be done by following the procedure in \cite{Series2} and constructing the permutationall invariant Bell operator corresponding to the expression (\ref{expression6}). Notice that to do so we assume for simplicity that each party performs the same measurements 
\begin{equation}
M_{0}^{(i)}  = \cos(\theta) \sigma_z^{(i)} +  \sin(\theta) \sigma_x^{(i)},\qquad
M_{1}^{(i)}  = \cos(\phi) \sigma_z^{(i)} +  \sin(\phi) \sigma_x^{(i)},
\end{equation}
where $\theta,\phi\in[0,2\pi)$, and $\sigma_z$ and $\sigma_x$ are the standard Pauli matrices.
Then, by computing the minimal eigenvalue of the resulting Bell operator $\mathcal{B} (\theta, \phi)$ and optimizing over the choice of angles, one obtains the maximal quantum violation of (\ref{expression6}) attainable with same measurements settings on each site.

By performing these numerical checks, it is possible to show that the bound $\beta_C^{k}$ for $k \leq 3 $ starts being violated for $N = 5$ parties, while for the higher cases $k = 4,5$ the violation appears from $N = 9$ and $N = 11$ respectively.
Moreover, if we take into account the analytical class of states introduced in Ref. \cite{Series2} (cf. Section 5.2), we can show that for a high enough number of parties it violates all the bounds that we have just derived. Indeed, let us recall that this class of states can achieve a relative violation $( \beta_Q - \beta_C^1) / \beta_C^1$ of  (\ref{expression6}) that tends to $-1/4$ when $N \rightarrow \infty$.
By using this result, it is easy to show that $\beta_Q$ exceeds $\beta_C^{k}$ for any $k \leq 6$, at least in the asymptotic limit.

To conclude, in the following Section we also present how to apply our results to an experimental setting.

\subsubsection{Experimental witnessing of $k$-body Bell correlations}
As shown in \cite{ScienceOur, Schmied, BaselMany} from permutationally invariant Bell inequalities one can derive Bell correlation witnesses. These are experimentally convenient inequalities allowing to reveal the presence of Bell correlations under the additional assumption that the measurement performed are known and correctly calibrated, as discussed earlier. In particular, the witness derived from Eq. (\ref{expression6}) has allowed the detection of Bell correlations in a Bose-Einstein condensate of $N\approx 500$ atoms \cite{Schmied} and in a thermal ensemble of $5\cdot 10^5$ atoms \cite{Engelsen}. 

In the same spirit, we show here how to derive a witness for Bell correlations of depth $k$ from the expression $\mathcal{I} + \beta_C^{k} \geq 0$, where $\mathcal{I}$ is defined in Eq. (\ref{expression6}). We assume that $M_{0}^{(i)}$ and $M_{1}^{(i)}$ are spin projection measurements on the $i$th party, along directions $\mathbf{n}$ and $\mathbf{m}$, respectively. This allows us to write $M_{0}^{(i)}=\boldsymbol{\sigma}^{(i)} \cdot \mathbf{n}$ and $M_{1}^{(i)}=\boldsymbol{\sigma}^{(i)} \cdot \mathbf{m}$, where $\boldsymbol{\sigma}^{(i)}$ is the vector of Pauli matrices acting on the $i$th party, and to express all correlators appearing in the Bell inequality as measurements of the collective spin operator $S_{\mathbf{n}} = (1/2)\sum_{i=1}^N \boldsymbol{\sigma}^{(i)}\cdot\mathbf{n}$. With the substitution $\mathbf{m}=2(\mathbf{a}\cdot\mathbf{n})\mathbf{a}-\mathbf{n}$ we arrive (see Ref. \cite{Schmied} for details) at the inequality
\begin{equation}\label{eq:kWitness}
- \left\vert \left\langle \dfrac{S_{\mathbf{n}}}{N/2} \right\rangle \right\vert + \left( \mathbf{a}\cdot\mathbf{n} \right)^2 \left\langle \dfrac{S_{\mathbf{a}}^2}{N/4} \right\rangle - \left( \mathbf{a}\cdot\mathbf{n} \right)^2 + \dfrac{\beta_C^{k}}{2N} \geq 0 \;,
\end{equation}
which is satisfied by all states with Bell correlations of depth at most $k$. In other words, the violation of Ineq.~\eqref{eq:kWitness} witnesses that the state of the system contains Bell correlations of depth (at least) $(k+1)$.

It is now convenient to define the spin contrast $\mathcal{C}_{\mathbf{n}}=\avg{2\op{S}_{\mathbf{n}}/N}$ and the scaled second moment $\zeta_{\mathbf{a}}^2 = \avg{4\op{S}_{\mathbf{a}}^2/N}$. Furthermore, we express $\mathbf{n} = \mathbf{a} \cos(\theta) + \mathbf{b} \sin(\theta)\cos(\phi) + \mathbf{c} \sin(\theta)\sin(\phi)$, with the ortho-normal vectors $\mathbf{a}$, $\mathbf{b}$ and $\mathbf{c}=\mathbf{a}\times\mathbf{b}$ with $\times$ denoting the vector product. With these definitions, we write Ineq.~\eqref{eq:kWitness} as
\begin{equation}\label{eq:kWitnessThetaPhi}
\zetaA^2 \geq \dfrac{\CA \cos(\theta) + \CB \sin(\theta)\cos(\phi) + \CC \sin(\theta)\sin(\phi) - \betaCk/(2N) + \cos^2(\theta)}{\cos^2(\theta)} \;,
\end{equation}
which is satisfied by states with Bell correlation depth at most $k$, for all $(\theta,\phi)$. For this reason, such states satisfy also
\begin{align}
\zetaA^2 &\geq Z_k(\CBC, \CA) = \max_{\theta\in[0,\pi]} \left[\dfrac{\CBC \sin(\theta) - \CA \cos(\theta) - \betaCk/(2N) + \cos^2(\theta)}{\cos^2(\theta)}\right]	\nonumber\\
&\geq Z_k(\CB,0) = \dfrac{2-\betaCk/(2N)-\sqrt{[\betaCk/(2N)]^2-\CB^2}}{2} \;,
\end{align}
where $\CBC=\sqrt{\CB^2+\CC^2}$ and where we used the monotonicity of the function $Z_k(\CBC,\CA)$, which is discussed in more detail in Ref. \cite{Schmied}. The inequality we obtain,
\begin{equation}\label{eq:kWitnessAB}
\zetaA^2 \geq \dfrac{2-\betaCk/(2N)-\sqrt{[\betaCk/(2N)]^2-\CB^2}}{2} \;,
\end{equation}
involves the measurements of $\zetaA$ and $\CB$, for the two orthogonal directions $\mathbf{a}$ and $\mathbf{b}$. The violation of Ineq.\eqref{eq:kWitnessAB}, for a given $\betaCk$, witnesses that the state contains Bell correlations with a depth of (at least) $k+1$.

\begin{figure}[!ht]
	\begin{center}
		\includegraphics[width=0.6\textwidth]{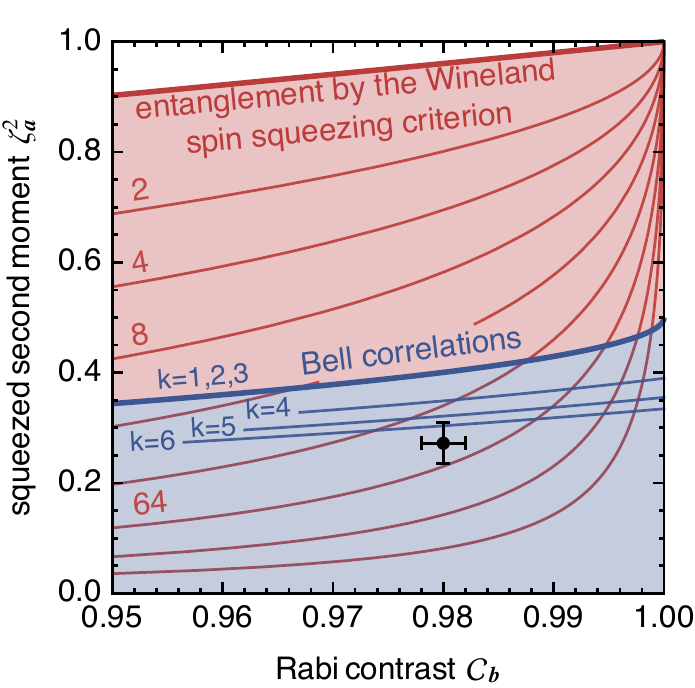}
		\caption{\label{fig:zetaviolation}
		\textbf{Quantification of the Bell correlation depth in a BEC with inequality~(\ref{eq:kWitnessAB}) and connection to spin squeezing and entanglement.}
		Black: the data reported in Ref. \cite{Schmied} expressed in terms of the Rabi contrast $\CB$ and the squeezed second moment $\zetaA^2$, with $1\sigma$ error bars. The number of particles is $N=480$.
		Blue shaded region: Bell correlations detected by violation of inequality~(\ref{eq:kWitnessAB}) for $k=1$.
		Red shaded region: entanglement witnessed by spin squeezing \cite{WinelandPRA1994, SoerensenNATURE2001}.
		Red lines: limits on $\zetaA^2$ below which there is at least $(k+1)$-particle entanglement \cite{SoerensenPRL2001}, increasing in powers of two up to $k=256$.
		Blue lines: limits on $\zetaA^2$ below which there are Bell correlations of depth at least $k+1$, for $k=1, \ldots, 6$.}
	\end{center}
\end{figure}

An interesting comparison is made with the Wineland spin-squeezing criterion \cite{WinelandPRA1994}, according to which entanglement is present if $\zetaA^2 < \CB^2$ \cite{Schmied}. This criterion was also shown to be able to quantify the degree of entanglement in the state \cite{SoerensenPRL2001}, $(k+1)$-particle entanglement is witnessed by measuring values of $\zetaA^2$ below some threshold, see Fig.~\ref{fig:zetaviolation} (red lines).

In Fig.~\ref{fig:zetaviolation} we plot the bounds given by Eq.~\eqref{eq:kWitnessAB}, for $k=1, \ldots, 6$, together with the entanglement bound obtained from the Wineland criterion \cite{SoerensenPRL2001}, and the experimental point measured in Ref. \cite{Schmied}. A statistical analysis on the probability distribution estimated experimentally \cite{Schmied} gives likelihoods of $99.9\%$, $97.5\%$, $90.3\%$ and $80.8\%$ for $1/2/3$-, $4$-, $5$- and $6$-body nonlocality respectively. This likelihood can be interpreted as, for example, a $p$-value of $1-80.8\%=19.2\%$ for rejecting the hypothesis: \textit{The experimental data were generated by a state that has no 6-body nonlocality, in the presence of Gaussian noise}.

\section{Conclusion}
\label{Conclusion}

We study the problem of detecting the nonlocality depth of quantum correlations. Nonlocality depth is a relevant concept in the study of multipartite systems, because it contains the information of how many particles share genuine Bell correlations in their state. In analogy to the case of nonlocality, detecting nonlocality depth is a computationally very demanding problem. Focusing on the property of these states we first derive a witness for genuine Bell correlations from known Svetlichny inequalities. We showed that this witness can reveal genuine Bell correlations with two collective measurements in systems where many-body correlation functions can be evaluated. We then adopt the framework of two-body symmetric correlations introduced in \cite{ScienceOur}, to meet the requirements of every-day experiments in many-body physics.

By developing a general framework to decribe the set of correlations of a given nonlocality depth, we are able to show that two-body symmetric correlations are enough to distinguish such depth. We do so by completely characterising the set of Bell inequalities that detect $k-$nonlocality with respect to nonsignaling resources for values of $k \leq 6$ and a fixed number of particle $N \leq 12$. Moreover, we take an explicit example of inequality and show that it can be used to witness the depth of Bell correlations for any number of parties. Lastly, we comment on the practical application of such a witness, by showing that its value can be evaluated using two collective measurements. We also illustrate how it can be successfully applied to already available experimental data from a Bose-Einstein condensate. 

Our results pave the way to a more refined study of Bell correlations in many-body systems, by presenting the first available techniques to determine the amount of particles sharing Bell correlations in these systems.
As a future direction to investigate, it would be interesting to derive inequalities that test for higher nonlocality depth than $6$, as it is already possible to do in the case of entanglement. In particular, a more ambitious direction would be to find ways to assess genuine Bell correlations in systems of hundreds of particles without relying on parity measurements. This would give a convenient way to prove that all the particles in the system are genuinely sharing Bell correlations.

As it is argued in the previous sections, the main challenge for these purposes consists in characterising the no-signalling set of multipartite correlations in the subspase of two-body permutationally invariant correlators.
We are able to do so only for the cases of low number of parties, while a general and efficient method is still missing. Therefore, a more technical but still interesting question would be to find such a general characterisation. 

Another possible research direction would be to consider inequalities involving more than two settings per party. The resulting witness could still only involve two measurement directions and provide improved bounds. In particular, it would be interesting to find the $k$-nonlocality bounds for the family of inequalities in~\cite{BaselMany} admitting an arbitrary number of settings.

It would also be interesting to design witnesses suited for specific families of states other than the GHZ state, such as the Dicke states or other graph states.

Except for the witnesses we derived from the Svetlichny inequalities, the witnesses we obtained here rely on the notion of Svetlichny models defined in terms of no-signalling resources. It would be interesting to see if the bounds derived here remain valid with respect to the sequential and signaling models, in which case they could also demonstrate this stronger form of $k$-nonlocality. Otherwise, it would be interesting to find other two-body inequalities suited for this task.

Lastly, we stress that our results can already be applied to experimentally detect in a Bell test genuine multipartite nonlocality for systems of size up to $N = 7$. In particular, since the inequalities that we introduce consist only of two-body correlators, such detection would require only an $\mathcal{O}(N^2)$ amount of mesasurements, contrarily to already known inequalities, such as Mermin's, that involve measuring an exponential amount of correlators.

\section{Acknowledgments}
We acknowledge support from the Spanish MINECO (SEVERO OCHOA Grant SEV-2015-0522, FISICATEAMO
FIS2016-79508-P, QIBEQI FIS2016-80773-P), ERC AdG OSYRIS (ERC-2013-ADG No. 339106) and CoG QITBOX (ERC-2013-COG No. 617337), 
the AXA Chair in Quantum Information Science, Generalitat de Catalunya
(2014-SGR-874, 2014-SGR-875 and CERCA Program), EU FET-PRO QUIC and Fundacio Privada Cellex, the
Swiss National Science Foundation (SNSF) through the NCCR QIST and the grant number
PP00P2-150579, the Army Research Laboratory Center for Distributed Quantum Information
via the project SciNet.
This project has received funding from the European Union's Horizon 2020 research and innovation programme under the Marie Sk{\l}odowska-Curie grant agreements No 748549 and No 705109.
M. F. was supported by the Swiss National Science Foundation through Grant No 200020\_169591.
\appendix

\section{Polytopes}
\label{AppA}

Here we provide the definition of polytopes and their duals. We then show how to project a polytope $P$ onto a given subspace $V$ as well as how to intersect $P$ and $V$.

A convex polytope, in what follows simply polytope, is a subset in a linear space $\mathbbm{R}^d$ with some finite $d$ defined as the convex hull of a finite number of points $\vec{e}_i\in \mathbbm{R}^d$ $(i=1,\ldots,K)$, i.e., 
\begin{equation}
P = \left\{ \vec{p} \in \mathbb{R}^d\, |\,  \vec{p} = \sum_{i = 1}^{m} q_i \vec{e}_i \, \,\, \mathrm{s.t.} \,\, \, q_i \geq 0 \,\, \, \mathrm{and} \,\, \, \sum_{i = 1}^m q_i = 1 \right\}.
\end{equation}
Here $\vec{e}_i$ are the vertices of $P$. Alternatively, one can define a polytope to be an intersection of a finite number of half-spaces, meaning that $P$ is described by a finite set of inequalities:
\begin{equation}\label{pol}
P = \left\{ \vec{p} \in \mathbb{R}^d\, |\,  \vec{f_i}\cdot\vec{p}\leq \beta_i \, \,\, \mathrm{with} \,\,\, \vec{f}_i\in\mathbbm{R}^d\,\,\,\mathrm{and}\,\,\, \beta_i\in\mathbbm{R} \right\}.
\end{equation}
Facets of a polytope $P$ are its intersections with the hyperplanes $\vec{f}_i\cdot \vec{p}=\beta_i$ is the facet of $P$. Let us also define the dual of a polytope $P$ to be 
\begin{equation}\label{dualpol}
P^{\ast} = \left\{ \vec{f}  \in \mathbb{R}^d\, |\,  \vec{f} \cdot \vec{p} \geq 0 \, \, \,\mathrm{for}\,\mathrm{all}\,\,\, \vec{p} \in P \right\}.
\end{equation}
For our convenience in Eqs. (\ref{pol}) and (\ref{dualpol}) we use different conventions regarding inequalities when defining a polytope and its dual.
However, it should be noticed that it is not difficult to transform inequalities appearing in Eq. (\ref{pol}) into those in Eq. (\ref{dualpol}) and \textit{vice versa}: in particular, to obtain inequalities in (\ref{dualpol}) from those in (\ref{pol}) it suffices to incorporate the free parameter $\beta_i$ and the sign into the vector $\vec{f}$, in the first case exploiting the fact that for a given choice of measurements, elements of the vector $\vec{p}$ are normalized.

Let us now discuss projections of polytopes. Consider a subspace $V\subset \mathbbm{R}^d$ and denote
by $\pi_V: \mathbbm{R}^d\to V$ the projection onto it. Imagine then that we want to project a given polytope $P\subset \mathbbm{R}^d$ onto $V$. There are two ways of determining the action of $\pi_V$ on $P$. First, $\pi_V(P)$ can be straightforwardly defined in terms of projection of its vertices, i.e., 
\begin{equation}\label{projection1}
\pi_V(P)=\left\{\vec{p}\in V\,|\, \vec{p}=\sum_{i=1}^{K}q_i\pi_V(\vec{e}_i)\,\,\,\mathrm{s.t.}\,\,\, q_i\geq 0\,\,\, \mathrm{and}\,\,\, \sum_{i}q_i=1\right\}.
\end{equation}
Notice that the projections of the vertices $\vec{e}_i$, $\pi_V(\vec{e}_i)$ might not be vertices of the projected polytope $\pi_V(P)$. On the other hand, a vertex of $\pi_V(P)$
must come from a vertex of $P$ under the projection $\pi_V$.

However, in certain situations it is much easier to describe a polytope by using inequalities instead of vertices. In fact, there are polytopes such as those formed by correlations fulfilling the no-signaling principle (see below) whose vertices are basically unknown, whereas their facets are straightforward to describe. In such case it is thus impossible to use (\ref{projection1}) in order to find $\pi_V(P)$, and one needs to exploit facets of $P$ for that purpose. A method that does the job is the Fourier-Motzkin elimination method \cite{FMEM}, which allows one to find facets of $\pi_V(P)$ starting from facets of $P$. 

Let us now briefly describe this method starting from an illustrative bidimensional example.
\begin{figure}[t]
\centering
\includegraphics[scale=0.5]{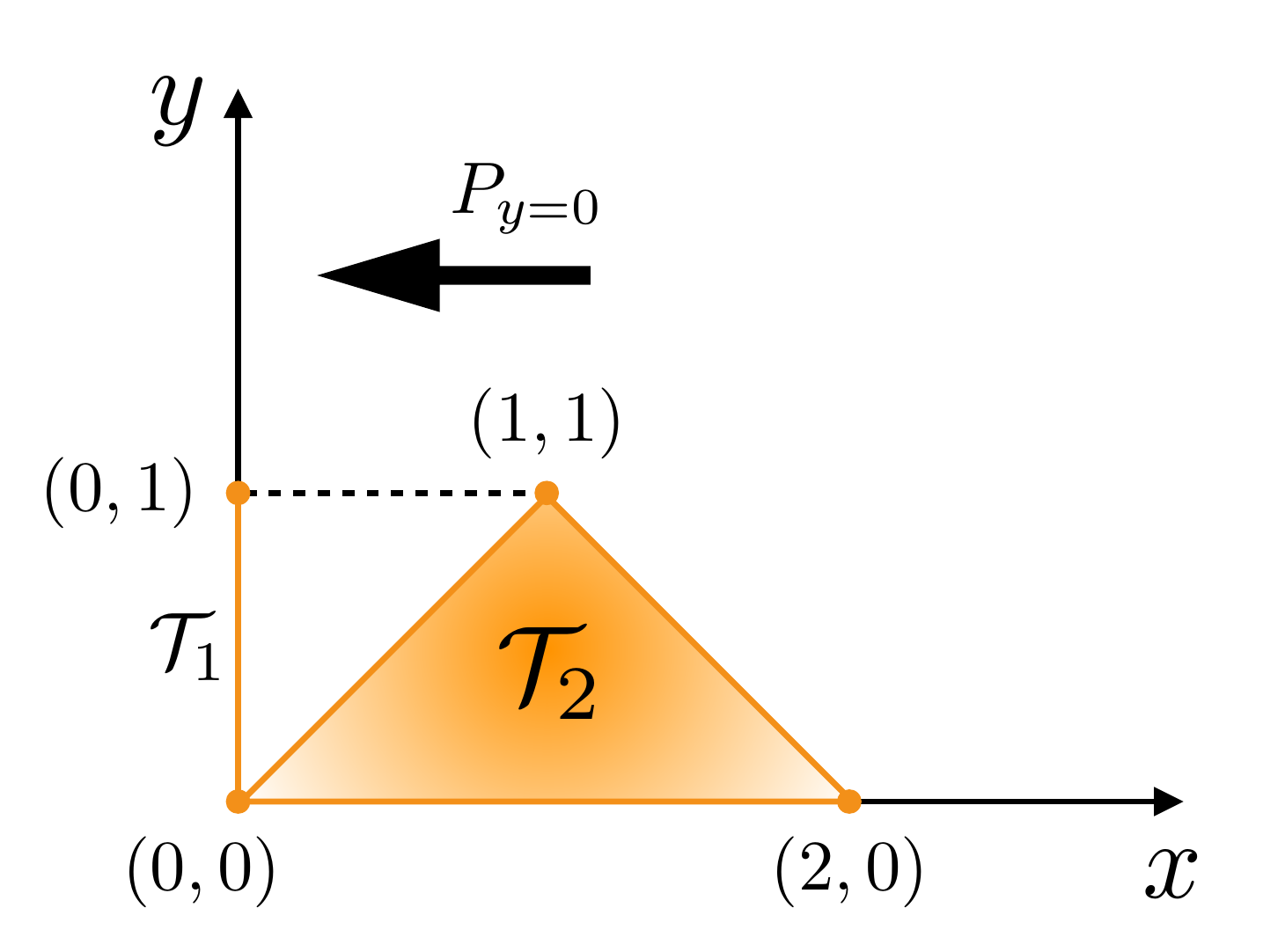}
\caption{Example of projection of a triangle onto the $y$ axis. The vertices of the triangle are the set $\lbrace (0,0), (2,0), (1,1) \rbrace$, whose projection are the points $\lbrace (0,0), (0,1) \rbrace$. Therefore the desired projected polytope is the set $0 \leq y \leq 1$.}
\label{triangle}
\end{figure}
Let us suppose that we want to project the triangle shown in Figure \ref{triangle} onto the $y$ axis. It is easy to see that the inequalities defining such geometrical object are
\begin{equation}\label{ineqTr}
x + y \leq 2,\quad
-x + y \leq 0, \quad y \geq 0.
\end{equation}
If we cancel out the $x$ coordinate from the inequalities, as we would have done when projecting vertices, we obtain the following three inequalities $y\leq 2$, $y\leq 0$
and $y\geq 0$,
which is not the projection we want since it defines only a single point $y = 0$. For further purposes, we notice that the result of this procedure coincides with the intersection of the polytope with the $x = 0$ axis, instead of the projection. Thus, projecting facets is a 
different task than projecting vertices: while for the latter it is enough to map 
each original vertex into the projected one, the above example shows that this procedure does not work for inequalities.

The basic principle of the Fourier-Motzkin elimination procedure is the fact that any convex combination of two facets of a polytope defines another valid inequality for it.
To be more precise, let us consider a polytope $P\subset\mathbbm{R}^{d}$ for some finite $d$, and let $\vec{f}_1 \cdot \vec{p} \leq \beta_1$ and $\vec{f}_2 \cdot \vec{p} \leq \beta_2$ be inequalities defining two different facets of it; here, $\vec{f}_1,\vec{f}_2\in \mathbbm{R}^d$ and $\beta_1,\beta_2 \in \mathbb{R}$, and $\vec{p}\in P$. It is clear that any vector $\vec{p}$ satisfying both these inequalities obeys also the following inequality
\begin{equation}
\left[ \lambda \vec{f}_1 + \left( 1 - \lambda \right) \vec{f}_2 \right] \cdot \vec{p} \leq  \lambda \beta_1 + \left( 1 - \lambda \right) \beta_2
\end{equation}
for any $0 \leq \lambda \leq 1$. The Fourier-Motzkin elimination exploit this property in order to define new valid inequalities bounding the polytope in which the coordinate that we want to project out is no longer involved. Coming back to the triangle example, we notice that by taking a convex combination with $\lambda = 1/2$ of the first two inequalities in (\ref{ineqTr}) we get a new inequality that involves only $y$, i.e., $y \leq 1$. If we consider in addition the third inequality, that does not contain $x$, we get the right projection of the triangle, that is the set $0 \leq y \leq 1$, as shown in Figure \ref{triangle}.

Let us now state the general procedure of the Fourier-Motzkin elimination. Given a generic polytope in $\mathbb{R}^d$ defined by a finite set of inequalities $\vec{f}_i \cdot \vec{p} \leq \beta_i$, where $\vec{f}_i\in \mathbbm{R}^d$ and $\beta_i\in\mathbbm{R}$, the list of inequalities defining its projection in the subspace defined by $p_{i} = 0$ for some $i$, is obtained through the following steps:
\begin{itemize}
\item divide the list of inequalities according to the sign of the coefficient in front of $p_{i}$ to obtain three sub-lists $f_{i_+}$, $f_{i_-}$,$f_{i_0}$ corresponding to positive, negative or zero coefficient,
\item take all the possible convex combinations between one element of $f_{i_+}$ and one of $f_{i_-}$, choosing the proper combination in order to get a new valid inequality with zero coefficient in front of $p_{i}$,
\item the obtained list, together with $f_{i_0}$, gives a complete set of inequalities that defines the projected polytope,
\item remove all the redundant inequalities to get the minimal set.

\end{itemize}

The main problem with the Fourier-Motzking elimination method is that it is in general very costly in terms of computational requirements. Indeed, due to the redundancy that one gets at each step, the time and memory needed to eliminate the variables scale exponentially with the number of variables that one wants to project out.

Another operation that we heavily exploit here is an intersection of a polytope $P$ with a given subspace. To define it let us consider again a linear space $\mathbbm{R}^d$ and its subspace $V\subset \mathbbm{R}^d$. Then, the intersection operation, denoted $\mathrm{int}_V : \mathbb{R}^d \longrightarrow V$ is defined as 
\begin{equation}
\mathrm{int}_V (P) =  \{ \vec{x} \in P \,|\, \vec{x} \cdot \vec{w} = 0 \, \, \,\mathrm{for}\,\mathrm{all} \,\,\, \vec{w} \in V^{\perp} \},
\end{equation}
where $V^{\perp}$ is the subspace of $\mathbbm{R}^d$ 
orthogonal to $V$. It is not difficult to notice that any elment belonging to the intersection of $P$ with $V$ is also an element of its projection onto the subspace, that is
\begin{equation}
\mathrm{int}_V (P) \subseteq \pi_V (P).
\label{incl}
\end{equation} 
Moreover, contrarily to the projection, the intersection of a polytope is more easily described in the dual representation. To show how, we define the dual basis $\lbrace \vec{v}_i^{\ast} \rbrace$ and $\lbrace \vec{w}_j^{\ast} \rbrace$ for the dual of the subspaces $V$ and $V^{\perp}$, respectively, so to decompose any inequality in $P^{\ast}$ as $\vec{f} = \sum_{i} f_i \vec{v}_i^{\ast} +  \sum_{j} f_j \vec{w}_j^{\ast} $. Then, we can define
\begin{equation}
\mathrm{int}_V (P)^{\ast} = \left\{ g \in V^{\ast}\, |\, g =  \sum_{i} f_i \vec{v}_i^{\ast} \, \, \, \mathrm{where}\,\,\, f_i = \vec{f} \cdot \vec{v}_i^{\ast} \, \, \, \mathrm{for} \, \, \vec{f} \in P^{\ast} \right\}.
\end{equation}
Before moving to the application to our specific case, we also notice that (\ref{incl}) implies, for the dual representation
\begin{equation}
\mathrm{int}_V (P)^{\ast} \supseteq \pi_V (P)^{\ast},
\label{incldual}
\end{equation}
meaning that some inequalities valid for the intersection of the polytope might be not valid for its projection. In other words, there are generally inequalities in $\mathrm{int}_V (P)^{\ast}$ that cannot be written as a convex combination of the original ones in $P^{\ast}$.

\section{Vertices of the projected nonsignaling polytopes for $k=2,3,4$}
\label{app:vertices}

Here we attach tables with vertices for the projections $\mathcal{NS}_{k}^{2,S}$ of the nonsignaling polytopes $\mathcal{NS}_k$ onto the symmetric two-body subspace for $2\leq k\leq 4$ (Tables \ref{table2}--\ref{table4}). For completeness we also attach the table
containing the deterministic values of single-body correlations (Table \ref{table1}). 
Notice that in the case $k=5,6$ the lists of vertices are too long to present it here. 
\begin{table}[h]
\centering
\begin{tabular}{|c||c c c c c|}
\hline
 & $S_0$ & $S_1$ & $S_{00}$ & $S_{01}$ & $S_{11}$ \\
\hline
$\xi_{1,1}$ & 1 & 1 & 0 & 0 & 0 \\[1ex]
$\xi_{1,2}$ & 1 & -1 & 0 & 0 & 0 \\[1ex]
$\xi_{1,3}$ & -1 & 1 & 0 & 0 & 0 \\[1ex]
$\xi_{1,4}$ & -1 & -1 & 0 & 0 & 0 \\
\hline
\end{tabular}
\caption{List of the values of the one and two-body symmetric expectation values for deterministic local strategies. In this case $S_0$ and $S_1$ contain consist of one expectation value, while $S_{mn}$ are simply zero.}
\label{table1}
\end{table}
\begin{table}[h]
\centering
\begin{tabular}{|c||c c c c c|}
\hline
 & $S_0$ & $S_0$ & $S_{00}$ & $S_{01}$ & $S_{11}$ \\
\hline
$\xi_{2,1}$ & 0 & 0 & 2 & 2 & -2 \\
$\xi_{2,1}$ & 0 & 0 & -2 & 2 & 2 \\
$\xi_{2,3}$ & 0 & 0 & 2 & -2 & -2 \\
$\xi_{2,4}$ & 0 & 0 & -2 & -2 & 2 \\
\hline
\end{tabular}
\caption{List of the vertices of $\mathcal{NS}_{2}^{2,S}$. In the first column we also add the corresponding populations.}
\label{table2}
\end{table}
\begin{table}[h]
\centering
\begin{tabular}{|c||c c c c c|}
\hline
 & $S_0$ & $S_0$ & $S_{00}$ & $S_{01}$ & $S_{11}$ \\
\hline
$\xi_{3,1}$ & -1 & -1 & 6 & -2 & -2 \\
$\xi_{3,2}$ & -1 & -1 & -2 & -2 & 6 \\
$\xi_{3,3}$ & -1 & 1 & 6 & 2 & -2 \\
$\xi_{3,4}$ & -1 & 1 & -2 & 2 & 6 \\
$\xi_{3,5}$ & 1 & -1 & 6 & 2 & -2 \\
$\xi_{3,6}$ & 1 & -1 & -2 & 2 & 6 \\
$\xi_{3,7}$ & 1 & 1 & 6 & -2 & -2 \\
$\xi_{3,8}$ & 1 & 1 & -2 & -2 & 6 \\
\hline
\end{tabular}
\caption{List of the vertices of $\mathcal{NS}_{3}^{2,S}$. In the first column we also add the corresponding populations.}
\label{table3}
\end{table}
\begin{table}[h]
\centering
\begin{tabular}{|c||c c c c c|}
\hline
 & $S_0$ & $S_0$ & $S_{00}$ & $S_{01}$ & $S_{11}$ \\
\hline
$\xi_{4,1}$ & -2 & -2 & 12 & 0 & 0 \\
$\xi_{4,2}$ & -2 & -2 & 0 & 0 & 12 \\
$\xi_{4,3}$ & -2 & 2 & 12 & 0 & 0 \\
$\xi_{4,4}$ & -2 & 2 & 0 & 0 & 12 \\
$\xi_{4,5}$ & 2 & -2 & 12 & 0 & 0 \\
$\xi_{4,6}$ & 2 & -2 & 0 & 0 & 12 \\
$\xi_{4,7}$ & 2 & 2 & 12 & 0 & 0 \\
$\xi_{4,8}$ & 2 & 2 & 0 & 0 & 12 \\
\hline
$\xi_{4,9}$ & 0 & 0 & 12 & -4 & -4 \\
$\xi_{4,10}$ & 0 & 0 & -4 & -4 & 12 \\
$\xi_{4,11}$ & 0 & 0 & 12 & 4 & -4 \\
$\xi_{4,12}$ & 0 & 0 & -4 & 4 & 12 \\
\hline
$\xi_{4,13}$ &$ \frac{-20}{7}$ & $ \frac{-4}{7}$ & $ \frac{36}{7}$ & $ \frac{-12}{7}$ & $ \frac{-12}{7}$ \\[1ex]
$\xi_{4,14}$ &$ \frac{-20}{7}$ & $ \frac{-4}{7}$ & $ \frac{-12}{7}$ & $ \frac{-12}{7}$ & $ \frac{36}{7}$ \\[1ex]
$\xi_{4,15}$ &$ \frac{-20}{7}$ & $ \frac{4}{7}$ & $ \frac{36}{7}$ & $ \frac{12}{7}$ & $ \frac{-12}{7}$ \\[1ex]
$\xi_{4,16}$ &$ \frac{-20}{7}$ & $ \frac{4}{7}$ & $ \frac{-12}{7}$ & $ \frac{12}{7}$ & $ \frac{36}{7}$ \\[1ex]
$\xi_{4,17}$ &$ \frac{20}{7}$ & $ \frac{-4}{7}$ & $ \frac{36}{7}$ & $ \frac{12}{7}$ & $ \frac{-12}{7}$ \\[1ex]
$\xi_{4,18}$ &$ \frac{20}{7}$ & $ \frac{-4}{7}$ & $ \frac{-12}{7}$ & $ \frac{12}{7}$ & $ \frac{36}{7}$ \\[1ex]
$\xi_{4,19}$ &$ \frac{20}{7}$ & $ \frac{4}{7}$ & $ \frac{36}{7}$ & $ \frac{-12}{7}$ & $ \frac{-12}{7}$ \\[1ex]
$\xi_{4,20}$ &$ \frac{20}{7}$ & $ \frac{4}{7}$ & $ \frac{-12}{7}$ & $ \frac{-12}{7}$ & $ \frac{36}{7}$ \\[1ex]
\hline
\end{tabular}
\caption{List of the vertices of $\mathcal{NS}_4^{2,S}$. In the first column we also present the associated populations.}
\label{table4}
\end{table}

\section{Complete list of facets for the polytopes that test for GMNL}\label{app:facets}

Here we present the complete list of facets for the polytopes that test for genuine multipartite nonlocality for $N = 3,4,5$. We omit the $N = 6,7$ cases since the amount of inequalities starts becoming too long to be contained in one page.
The inequalities are sorted in equivalence classes, under symmetry operations such as outcome/input swapping, in the same fashion as in \cite{ScienceOur}.
\begin{table}[h]
\centering
\begin{tabular}{|cccccc|} 
\hline
$\beta_C$ & $\alpha$ & $\beta$ & $\gamma$ & $\delta$ & $\epsilon$ \\
\hline
1 & 0 & 0 & 1 & 0 & 0\\ 
12 & -3 & 1 & 3 & - $\frac{3}{2} $& -2\\ 
6 & -2 & -2 & 0 & 1 & 0\\ 
3 & 0 & 0 & 0 & -1 & 1\\ 
3 & 0 & -2 & 0 & 0 & 1\\ 
3 & 0 & 0 & -1 & 0 & 0 \\
\hline
\end{tabular}
\caption{List of the facets of the symmetric two-body polytope of 2-producible correlations for $N= 3$}
\label{facets3}
\end{table}

\begin{table}[h]
\centering
\begin{tabular}{|cccccc|} 
\hline
$\beta_C$ & $\alpha$ & $\beta$ & $\gamma$ & $\delta$ & $\epsilon$ \\
\hline
2 & 1 & 0 & 1 & 0 & 0\\ 
42 & 12 & 3 & 6 & 2 & -3\\
 42 & -12 & 9 & 6 & -6 & 1\\
  20 & -5 & 3 & 4 & -3 & 0\\ 
  30 & -6 & 3 & 6 & -4 & -1\\ 
  12 & 0 & 0 & 3 & 1 & -1\\ 
  12 & 3 & 3 & 1 & 2 & 1\\ 
  6 & -3 & 0 & 1 & 0 & 0\\ 
  8 & -3 & -1 & 2 & 1 & 0\\ 
  6 & 0 & 0 & 1 & -1 & 0\\ 
  8 & 0 & 2 & 1 & 1 & 1\\ 
  12 & -3 & -3 & 0 & 1 & 0\\ 
  6 & 0 & 0 & -1 & 0 & 0 \\
\hline
\end{tabular}
\caption{List of the facets of the symmetric two-body polytope of 3-producible correlations for $N= 4$}
\label{facets4}
\end{table}

\begin{table}[h]
\centering
\begin{tabular}{|cccccc|} 
\hline
$\beta_C$ & $\alpha$ & $\beta$ & $\gamma$ & $\delta$ & $\epsilon$ \\
\hline
30 & -4 & 10 & 1 & -2 & 3\\ 
40 & -8 & 12 & 1 & -3 & 3\\ 
116 & 28 & -28 & 4 & -9 & 4\\ 
134 & -36 & 30 & 8 & -11 & 4\\ 
452 & -120 & 104 & 25 & -37 & 15\\ 
562 & -144 & 136 & 27 & -45 & 22\\ 
112 & 28 & -28 & 5 & -9 & 5\\ 
116 & 28 & -28 & 5 & -10 & 5\\ 
380 & -92 & 84 & 21 & -34 & 13\\ 
36 & -4 & 8 & 4 & -3 & 2\\ 
380 & -92 & 84 & 20 & -33 & 12\\ 
320 & -76 & 68 & 20 & -29 & 10\\ 
200 & -52 & 44 & 12 & -17 & 6\\ 
16 & -2 & 4 & 2 & -1 & 1\\ 
110 & -30 & 24 & 7 & -9 & 3\\ 
20 & 4 & -4 & 0 & -1 & 0\\ 
410 & -120 & 72 & 40 & -30 & 3\\ 
170 & -60 & 24 & 20 & -10 & 1\\ 
8 & 0 & 0 & 2 & 1 & 0\\ 
20 & 0 & 0 & 3 & 3 & 1\\ 
20 & -2 & 8 & 0 & -1 & 3\\ 
20 & 0 & 4 & 1 & -2 & 3\\ 
50 & 0 & 12 & 4 & -4 & 5\\ 
40 & -4 & 12 & 2 & -3 & 4\\ 
80 & 4 & 12 & 9 & -8 & 7\\ 
34 & 2 & 6 & 4 & -3 & 3\\ 
4 & 2 & 0 & 1 & 0 & 0\\ 
10 & -4 & 0 & 1 & 0 & 0\\ 
220 & 60 & 12 & 20 & 5 & -8\\ 
120 & 20 & -4 & 20 & -5 & -6\\ 
400 & -60 & 36 & 60 & -45 & 2\\ 
20 & 4 & 2 & 3 & 2 & 0\\ 
20 & -4 & -4 & 1 & 2 & 1\\ 
80 & 8 & 20 & -2 & 5 & 10\\ 
40 & -12 & -6 & 5 & 3 & 0\\ 
2 & 0 & 0 & 0 & 0 & 1\\ 
10 & 0 & 0 & -1 & 0 & 0 \\
\hline
\end{tabular}
\caption{List of the facets of the symmetric two-body polytope of 4-producible correlations for $N= 5$}
\label{facets5}
\end{table}

\end{document}